\documentclass[]{aa} % for the letters 
\usepackage{graphicx}
\usepackage{txfonts}
\usepackage{mathabx}
\usepackage{hyperref}
\hypersetup{
    colorlinks=true,
    linkcolor=black,%blue,
    filecolor=black, %blue,
    urlcolor=black, %blue,
    citecolor=black, %blue,
}

\usepackage{multirow}
\usepackage{amsmath,array,graphicx}
\usepackage{kantlipsum}
\usepackage{comment}
\newcommand{\orcid}[1]{\href{https://orcid.org/#1}{\includegraphics[width=8pt]{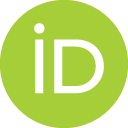}}}
\begin{document} 
 \bibliographystyle{aa}

   \title{First Comparative Exoplanetology Within a Transiting Multi-planet System: Comparing the atmospheres of V1298 Tau b and c }

   %\subtitle{}

   \author{Saugata Barat
          \inst{1}\orcid{0009-0000-6113-0157}
          \and
          Jean-Michel D\'esert
          \inst{1}\orcid{0000-0002-0875-8401}
          \and Jayesh M. Goyal
          \inst{2} \orcid{0000-0002-8515-7204}
          \and
          Allona Vazan \inst{3} \orcid{0000-0001-9504-3174}
          \and
          Yui Kawashima \inst{4,5,6,7} \orcid{0000-0003-3800-7518}
          \and
          Jonathan J. Fortney \inst{8} \orcid{0000-0002-9843-4354}
          \and
          Jacob L. Bean \inst{9} \orcid{0000-0003-4733-6532}
          \and
          Michael R. Line \inst{10} 
          \and
          Vatsal Panwar \inst{11,12} \orcid{0000-0002-2513-4465}
          \and
          Bob Jacobs \inst{1} \orcid{0000-0002-0373-1517}
          \and
          Hinna Shivkumar \inst{1}
          \and
          James Sikora \inst{1,13} \orcid{0000-0002-3522-5846}
          \and 
          Robin Baeyens \inst{1}  
          \and
          Antonija Oklop{\v{c}}i\'c \inst{1} \orcid{0000-0002-9584-6476}
          \and
          Trevor J. David \inst{14} \orcid{0000-0001-6534-6246}
          \and
          John H. Livingston \inst{15,16,17} \orcid{0000-0002-4881-3620}
          }

   \institute{Anton Pannekoek Institute for Astronomy, University of Amsterdam,
Science Park 904, 1098 XH,
Amsterdam, the Netherlands \\
    \email{s.barat@uva.nl}
    \and 
    School of Earth and Planetary Sciences (SEPS), National Institute of Science Education and Research (NISER), Jatani, India
    \and
    Astrophysics Research Center (ARCO), The Open University of
Israel, Ra’anana, 43107, Israel
\and
Frontier Research Institute for Interdisciplinary Sciences, Tohoku University, 6-3 Aramaki aza Aoba, Aoba-ku, Sendai, Miyagi 980-8578, Japan
\and
Department of Geophysics, Graduate School of Science, Tohoku University, 6-3 Aramaki aza Aoba, Aoba-ku, Sendai, Miyagi 980-8578, Japan
\and
Institute of Space and Astronautical Science, Japan Aerospace Exploration Agency, 3-1-1 Yoshinodai, Chuo-ku, Sagamihara, Kanagawa 252-5210, Japan
\and
Cluster for Pioneering Research, RIKEN, 2-1 Hirosawa, Wako, Saitama 351-0198, Japan
\and
Department of Astronomy \& Astrophysics, University of
California, Santa Cruz, CA 95064, USA.
\and
Department of Astronomy and Astrophysics, University of
Chicago, Chicago, IL, USA
\and
School of Earth and Space Exploration, Arizona State
University, Tempe, AZ 85287, USA
\and
Department of Physics, University of Warwick, Coventry CV4
7AL, UK
\and
Centre for Exoplanets and Habitability, University of Warwick,
Coventry CV4 7AL, UK.
\and
Lowell Observatory, 1400 W Mars Hill Road, Flagstaff, AZ, 86001, USA
\and
Center for Computational Astrophysics, Flatiron Institute, New
York, NY 10010, USA.
\and
Astrobiology Center, 2-21-1 Osawa, Mitaka, Tokyo 181-8588,
Japan.
\and
National Astronomical Observatory of Japan, 2-21-1 Osawa,
Mitaka, Tokyo 181-8588, Japan.
\and
Department of Astronomical Science, The Graduate University for Advanced Studies, SOKENDAI, 2-21-1 Osawa, Mitaka, Tokyo 181-8588, Japan.}

   \date{}

% \abstract{}{}{}{}{} 
% 5 {} token are mandatory
 
  \abstract
   { 

 The V1298 Tau system is a multi-planet system that provides the opportunity to perform comparative exoplanetology between planets orbiting the same star. Because of its young age (20-30~Myr), this system also provides the opportunity to compare the planet's early evolutionary properties, right after their formation. 
 We present the first atmospheric comparison between two transiting exoplanets within the same multiple planet system: V1298 Tau b  and V1298 Tau c. We observed one primary transit for each planet with the Hubble Space Telescope (HST), using Grism 141 (G141) of Wide Field Camera 3 (WFC3). We fit the spectroscopic light curves using state-of-the-art techniques to derive the transmission spectrum for planet c and adopt the transmission spectrum of planet b obtained with the same observing configuration and data analysis methods from previous studies.  
We measure the mass of planet b and c ($8_{-2}^{+4}$, $17_{-6}^{+13}$~M$_{\oplus}$; respectively) from the transmission spectrum and find the two planets to have masses in the Neptune/sub-Neptune regime.  
 Using atmospheric retrievals, we measure and compare the atmospheric metallicities of planet b and c (logZ/Z$_\odot$=-2.04$_{-0.59}^{0.69}$, logZ/Z$_\odot$= -0.16$_{-0.94}^{1.15}$, respectively), and find them consistent with solar/sub-solar, which is low (at least 1 order or magnitude) compared to known mature Neptune/sub-Neptune planets. 
 This discrepancy could be explained by ongoing early evolutionary mechanisms, which are expected to enrich the atmospheres of such young planets as they mature. Alternatively, the observed spectrum of planet c can be explained by atmospheric hazes, which is in contrast to planet b, where efficient haze formation can be ruled out. Higher haze formation efficiency in planet c could be due to differences in atmospheric composition, temperature and higher UV flux compared to planet b. In addition,  planet c is likely to experience a  higher fraction of mass loss compared to planet b, given its proximity to the host star. }

 \titlerunning{Comparative exoplanetology between V1298 Tau b and V1298 Tau c}
\authorrunning{Saugata Barat}
   \maketitle
%
%-------------------------------------------------------------------

\section{Introduction}

Demographic studies on mature exoplanets \citep{szabo2011,fulton2017,fulton2018,van_eylen} suggest that early evolutionary processes like photoevaporation \citep{owen2017} and core-powered mass loss \citep{ginzburg2018,gupta2019} significantly influence the structure and composition of low mass planets, such as Neptunes and sub-Neptunes. Therefore, observing planets right after their formation is the key to understanding these early evolutionary processes.

V1298 Tau, a 20-30~Myr old weak-lined T-Tauri star in the foreground of the Taurus-Auriga star-forming region \citep{oh17,luhman18}, is particularly valuable for such studies. It has a temperature of $\sim$5000~K, with a radius and mass of 1.34$\pm$0.05~R$_{\odot}$ and 1.1$\pm$0.05~M$_{\odot}$, respectively, and has a spectral type between K0-K1 \citep{david2019}.  It is a pre-main sequence star, and shows rotational variability at the level of $\sim$2\% peak-to-peak and a rotational period of 2.86$\pm$0.01 days \citep{david2019,Feinstein2022,sikora2023}. Such rotational variability is typical of young pre-main sequence stars \citep[e.g. see][]{david2016,mann2016,mann2022} and is due to the presence of spots and faculae on their photospheres. \citet{feinstein2021} reproduced the optical photometric variability using a model with 20\% spot coverage on the surface assuming a temperature contrast of 500~K between the photosphere and the spots.

V1298 Tau hosts four transiting planets, making it one of the youngest known transiting multi-planet systems. The inner three planets (V1298 Tau c, d, and b) are in a near 3:2:1 mean-motion resonance \citep{david2019b,Feinstein2022,sikora2023}, suggesting possible migration within a gas disk \citep{terquem2018,kajtazi2023}. The mutual inclinations between the planets in this system are known to be low \citep{marshall2021}, indicating planar orbital geometry for this system. The orbital period of the outermost planet, V1298 Tau e, remains unconfirmed but is under investigation \citep{damiano2023}.

With a wealth of observations of exoplanet atmospheres, we have entered an era of understanding atmospheric processes, planet formation and early evolution through comparative studies. There have been previous studies aimed at understanding the physical and chemical processes of mature transiting exoplanet atmospheres from a statistical point of view. For example, the dependence of thermal structure and opacity sources in the atmosphere on insolation \citep[e.g, see][]{garhart2020,baxter2020a},  emergence of clouds in the atmosphere \citep{sing2016,Fu2017,keating2019}, disequilibrium chemistry \citep{Baxter2021,fortney_2020,tsai2023,baeyens2022} and the formation and composition of hazes \citep{gao2020b,Brande2024} has been studied by comparing the atmospheres of different exoplanets. But, differences in the host stars can introduce biases due to uncertainty in their elemental abundances, activity levels and evolution history. These challenges can be mitigated by performing comparative exoplanetology between multiple planets in the same system, since they share their host star. In this paper, we compare the atmospheric metallicity, chemistry and potential evolution of two planets in the V1298 Tau system: V1298 Tau b (0.9~R$_J$, 24.14 day period) and V1298 Tau c (0.5~R$_J$, 8.24 day period) by studying their atmospheres. Given its young age, this system also allows for atmospheric comparison with mature transiting exoplanets as well as, widely separated directly imaged planets which have similar age. A comparison between the known physical properties of planet b and c is shown in Table \ref{tab:b_c_comparison}.

 V1298 Tau c receives four times the stellar irradiation as V1298 Tau b due to their orbital distance ratio of $\sim$1:2. Stellar XUV flux is believed to play an important role in atmospheric chemistry \citep{venot2015,kawashima2018,shulyak2019,tsai2023} as well as early evolution \citep{owen2013,owen2017}, which we aim to test with two planets in the same young system.

Mass measurement for these young planets is challenging due to stellar activity induced jitter affecting radial velocity (RV) observations \citep{bremms2019,blunt2023}. Initial RV studies \citep{mascareno_2021} reported Jovian masses (200$\pm$70, 400$\pm$100~M$_{\oplus}$) for the two outer V1298 Tau planets (b and e, respectively) and mass upper limits (70 and 100~M$_{\oplus}$) for the inner two (c and d, respectively). Follow-up RV observations of this system have challenged these mass measurements \citep{sikora2023,finoceti2023}. \citet{finoceti2023} reports an upper limit of $\sim$50~M$_{\oplus}$ for planet c and d and an upper limit of $\sim$150~M$_{\oplus}$ for planet b. \citet{sikora2023} reports a mass of 20$\pm$10~M$_\oplus$ for planet c and upper limits of 36 and 160~M$_\oplus$ for planets d and b, respectively. The follow-up RV papers are consistent with \citet{mascareno_2021} for the mass of V1298 Tau e, but do not agree with the mass of planet b from \citet{mascareno_2021}.  \citet{barat2023} estimated V1298 Tau b's mass to be less than 23~M$_\oplus$ at 3$\sigma$ level of confidence from HST/WFC3 transmission spectra, and ruled out RV estimate from \citet{mascareno_2021} at 5$\sigma$ level of confidence. Ongoing Transit timing variations (TTVs) studies (Livingston et al in prep) suggest that both V1298 Tau b and c are in the sub-Neptune mass regime.

In this paper, we compare the transmission spectra of V1298 Tau b and c. For planet b, we adopt the spectrum obtained in \citet{barat2023}. We analyze the primary transit observations of V1298 Tau c, observed using the same instrumental configuration as \citet{barat2023}, and present the transmission spectrum of planet c. We discuss the data reduction and light curve analysis for the visit of V1298 Tau c in Section \ref{sec:data analysis}, followed by the main results in Section \ref{sec:results}. We discuss and interpret our findings in Section \ref{sec:discusison} and \ref{sec:conclusion}.

\begin{table}
    \centering
\caption{Comparison of the the physical properties of V1298 Tau b and c}
  \resizebox{\columnwidth}{!}{ \begin{tabular}{c|c|c}
        
        Parameter & V1298 Tau b & V1298 Tau c\\
         \hline
         \hline
      Period [days] \tablefootmark{1}   & 24.140410$\pm$0.000022  & 8.248720$\pm$0.000024 \\
      Semi-major axis [AU]   & 0.1716$\pm$0.0028 & 0.0839$\pm$0.0014\\
      Equilibrium temperature [K]   & 685$\pm$15  & 979$\pm$21\\
      Radius [R$_\oplus$]   & 9.95$\pm$0.37 & 5.24$\pm$0.24\\
 %     Ephemeris reference   & \citet{sikora2023} & \citet{sikora2023} \\
      Known mass [M$_\oplus$] & $<$23 \tablefootmark{2}   & 20$\pm$10 \tablefootmark{1} \\
  %    Mass reference & \citet{barat2023} & \citet{sikora2023} \\
         \hline
    \end{tabular}}
%    \caption{Comparison of the the physical properties of V1298 Tau b and c}
    \tablefoot{\tablefoottext{1}{\citet{sikora2023}} \tablefoottext{2}{\citet{barat2023}}}
    \label{tab:b_c_comparison}
\end{table}

%\vspace{-1cm}

\section{Data analysis and modeling} \label{sec:data analysis}

\subsection{Observations} \label{subsec:obs}
We observed one primary transit of V1298 Tau c (18th October 2021) with 8 HST orbits using the WFC3 instrument for Program GO 16462. The observations were taken with the G141 in bidirectional spatial scanning mode covering a range of 1.1-1.7~$\mu$m. We used the $256\times256$ pixel subarray and  \texttt{SPARS25}, \texttt{NSAMP}=5 readout mode, which resulted in 88.4~s exposures. We used a 0.23"/second scan rate, resulting in a 170 pixel long image in the spatial direction. The observations for V1298 Tau b had been taken using the same instrumental settings, with 10 HST orbits. See \citet{barat2023} for further details.

\subsection{Data reduction and light curve analysis} \label{subsec:data reduction}

We use a custom data reduction pipeline for analyzing the raw HST spatially scanned images \citep{Arcangeli2018a,arcangeli2019,jacobs2022,barat2023} for the V1298 Tau c visit. The same pipeline had been used for analyzing the visit of planet b in \citet{barat2023}. We outline the data reduction methods we use in Appendix \ref{appendix:pipleine}.

\begin{figure}

\centering
\includegraphics[width=\linewidth]{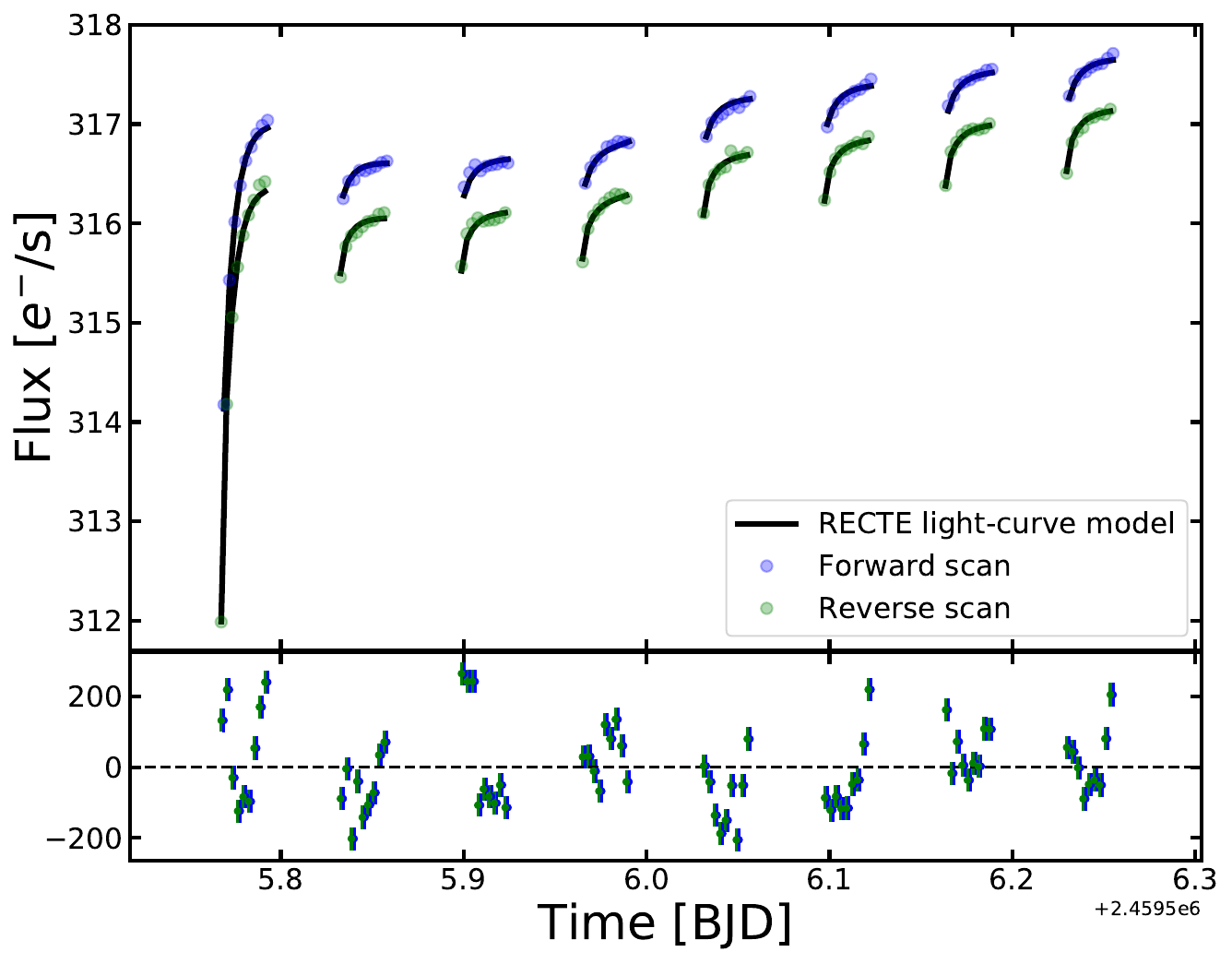}
%\label{fig:whitelc}
\caption{ Upper panel: Raw white (1.12-1.65$\mu$m band) light-curve for the transit of V1298 Tau c. Blue and green points show forward and reverse scanned exposures, respectively. Black solid lines show best fit transit model. We use a physically motivated charge trapping model, RECTE \citep{zhou2017,Zhou2020} to fit the exponential ramps. The baseline is modelled using  a quadratic function in time. See Appendix \ref{subsection:stellar baseline} for further discussion. The transit has been modelled using \texttt{batman} \citep{batman}. For details of light-curve fitting see Section \ref{sec:data analysis}. Lower panel: The residuals from the fit to the white light curve observations.}
\label{fig:whitelc}
\end{figure}

The broadband integrated white light curve for the visit of V1298 Tau c is shown in Figure \ref{fig:whitelc}. It shows `hook' like systematics for each orbit, which have been observed for similar HST observations \citep{berta2012,deming13,ranjan2014}. We also see that the orbital ramp is much larger in the first orbit compared to the rest of the orbits for this visit. This is a typical feature of HST visits where the first orbit shows larger systematics \citep{wakeford2016,arcangeli2021}. The light curves also show a long-term curvature, which is likely due to the rotational variability from active regions on the photosphere of this young star. This variability has been reported by \citet{barat2023} and is well known for young pre-main sequence stars which have large spot coverage and fast rotation rates \citep[e.g. see][]{david2016,david2019b,mann2022}. Such variability is known to affect multi-epoch transmission spectra \citep[e.g,][]{desert2011b,berta2012,rackham2019}.

These observations of planet c were planned using linear ephemeris from \citet{Feinstein2022}, but this system is known to exhibit TTVs and the actual transit occurred $\sim$2 hours prior to the predicted time from the linear ephemeris. Therefore, our observations covered only one pre-transit orbit and no exposures during ingress or egress. Typically, the first orbit of a visit for HST is known to exhibit large systematics, and it is common practice not to include it in time series analysis \citep[e.g. see][]{berta2012,ranjan2014,jacobs2022,barat2023}. In the case of V1298 Tau b, there were three pre-transit orbits, and therefore, the first orbit could be removed from the analysis. However, since we have only one pre-transit orbit for the visit of V1298 Tau c, we could not remove the first orbit while analyzing its light curve.

 The white light curve of planet c has been modeled assuming a physically motivated charge-trapping model RECTE \citep{zhou2017} to model the `hook-like' orbital ramps and a second order polynomial function to model the baseline for the planet c visit. The RECTE model is known to fit the first orbit better compared to other known methods \citep[e.g,][]{berta2012}. The planet transit was modeled using \texttt{batman} \citep{batman}, with orbital parameters taken from \citet{sikora2023}. This same approach had been used to fit the white light curve of planet b \citep{barat2023}. 

A linear limb darkening model was assumed for V1298 Tau. \citet{barat2023} fit for the linear limb darkening coefficient for the visit of planet b. Since the planet c visit does not have any ingress/egress coverage, we have used the fitted limb darkening for V1298 Tau from \citet{barat2023} as a fixed parameter for the analysis of the planet c visit (Figure \ref{fig:limb_darkening_comparison}).  V1298 Tau b and c have impact parameters (0.34$^{+0.19}_{-0.21}$ and 0.46$^{+0.13}_{-0.24}$,  respectively) consistent within 1$\sigma$ of each other \citep{david2019b,david2019}. Furthermore, \citet{marshall2021} report a relatively low mutual inclination (0$\pm$19$^\circ$), indicating a planar orbital geometry of the V1298 Tau system. Therefore, both planets would transit a similar part of the stellar photosphere, and are expected to to show similar limb darkening during their primary transits. Further details of the light curve fitting methods are provided in Appendix \ref{light curve fitting}.

 We obtain 17 spectroscopic light curves for the planet c visit by using 7-pixel bins. The same binning had been used for planet b in \citet{barat2023}.
For the analysis of spectroscopic light curves of planet c, the semi-major axis is fixed from the white light curve fit of planet c. This approach had also been followed for the planet b visit. In the analysis of planet b \citep{barat2023}, the common-mode method (using white light curve systematics models to correct the spectroscopic light curves) was applied to de-trend the spectroscopic light curves, assuming that the systematics were achromatic. However, the first orbit is known to exhibit wavelength-dependent systematics \citep{zhou2017,Zhou2020}, and therefore in the analysis of V1298 Tau c we could not use the common-mode technique for de-trending spectroscopic light curves. We used RECTE \citep{zhou2017} to model the `hook-like' orbital ramps and a second order polynomial function to model the baseline of the spectroscopic light curves for the planet c visit. The parameters for the RECTE as well as the polynomial coefficients, were fitted for each spectroscopic light curve. We tested different models for the baseline (linear, quadratic and cubic). We found that the quadratic model fits the observed baseline the best. See Appendix \ref{subsection:stellar baseline} for further details on the baseline model. The limb darkening coefficients were fixed from \citet{barat2023}.

We derive the transmission spectrum of V1298 Tau c following the approach outlined in this Section, and we present it in Figure \ref{fig:comparison_spectrum}, along with the spectra of planet b, which has been adopted from \citet{barat2023}.

 \subsection{Atmospheric models from ATMO} \label{atmo_model}

 We interpret the transmission spectrum of V1298 Tau c using the equilibrium chemistry ATMO retrieval framework \citep{Tremblin2015,Drummond2016, Goyal2018}. In this paper, for a consistent comparison with planet c, we also reanalyze the spectrum of V1298 Tau b adopted from \citet{barat2023}, using the same framework as applied for planet c.  ATMO has been previously used to analyze various exoplanet atmospheres \citep[see for e.g][]{Evans2018,Goyal2019,Carter2020}. ATMO's python based retrieval framework was applied and benchmarked in \citet{Lewis2020} and \citet{Rathcke2021} for HAT-P-41b and WASP-79b, respectively. This same retrieval framework is applied to V1298 Tau b and c in this work. This framework includes a Nested sampling Bayesian sampler DYNESTY \citep{Speagle2020} coupled to ATMO.
 
 We model the planets using a 1D atmosphere with an isothermal T-P profile and calculate the molecular abundances, assuming chemical equilibrium. Atmospheric metallicity and C/O ratio are used as parameters to calculate the elemental abundances. For a given metallicity, in our model the C/O ratio is implemented by keeping the number of C atoms fixed, and adjusting the number of O atoms. We include a grey opacity source to model clouds. We include  H$_2$-H$_2$, H$_2$-He Collision Induced opacity (CIA), as well as H$_2$O, CH$_4$ and CO high temperature molecular opacity in the retrieval. These molecules are chosen, as the HST bandpass includes absorption features from these molecules. We include planet mass as a free parameter in our retrieval setup for both planets b and c. We put similar uniform priors on atmospheric metallicity (0.01 - 1000$\times$ solar), O/H ratio (0.01 - 100$\times$ solar), degree of cloudiness (0.1 - 100$\times$ Rayleigh scattering cross-section at 350~nm) for both V1298  Tau b and c. We put uniform prior on the mass between 5 - 50~M$_{\oplus}$ for planet c \citep{finoceti2023,sikora2023} and between 5-25~M$_{\oplus}$ for planet b \citep{barat2023}.

\section{Results} \label{sec:results}

\subsection{The transmission spectrum of V1298 Tau c} \label{subsec:transmission spectrum}

We ran a free atmospheric chemistry retrieval using the open-source code \texttt{PetitRadtrans} \citep{molliere2019} on the transmission spectrum of planet c. We fixed the planet mass and temperature to 17~M$_{\oplus}$ and 950~K, respectively. These mass and temperature values for V1298 Tau c are taken from the best fit models to the spectrum from ATMO models, described in Section \ref{subsec:atmo}. We included a grey cloud deck opacity in the free retrieval. The posterior distribution from this fit is is shown in Figure \ref{fig:free corner}. Our models find water vapor in the atmosphere of planet c, with a lower limit for the water volume mixing ratio ($>10^{-5}$ at 3$\sigma$ confidence level). We do not find any constraints on the methane abundance. 

We test the robustness of the water absorption feature indicated by the free retrieval. We compared the observed transmission spectrum of V1298 Tau c (Figure \ref{fig:comparison_spectrum}) to a flat line using a Chi-square test. We found that models with atmospheric water absorption are slightly favored (2.5$\sigma$) by comparing the difference in chi-square values between a flat line and the best fit model generated using median values of the fitted parameters from the free retrieval.

 We also tested the impact of fixing the planet mass and isothermal temperature, by including Gaussian priors at 950~K and 17~M$_{\oplus}$ with widths of 10~K and 1~M$_{\oplus}$. The retrieved posterior distributions and values for the water and methane abundance agree within 1$\sigma$ to the ones derived from the fixed mass-temperature retrievals, shown in Figure \ref{fig:free corner}. Thus, we conclude that with different sets of assumptions, our retrievals suggest the presence of water vapour in the atmosphere of planet c.

The transit depth uncertainties for planet c are larger by a factor of about 1.5 compared to the spectrum of planet b (Figure \ref{fig:comparison_spectrum}), adopted from \citet{barat2023}. In Section \ref{subsec:data reduction}, we discuss the differences in the light curve analysis between planet c and b. Since, we cannot ignore the first orbit for the visit of planet c, unlike the case for planet b visit, we cannot apply a common-mode correction to the spectroscopic light curves of planet c. Therefore, the higher uncertainty in measured transit depth for V1298 Tau c is likely due to the modelling of each spectroscopic light curve independently using RECTE to model the charge trapping and second order polynomial to model the baseline. The common-mode correction could correct for the curvature of the time-series baseline, allowing a linear baseline model as shown in the case of planet b \citep{barat2023}.

\subsection{Results from ATMO equilibrium chemistry retrievals} \label{subsec:atmo}

In this Section, we summarize the results from the equilibrium chemistry atmospheric retrievals using ATMO for both planet b and c. The ATMO models are described in Section \ref{atmo_model}. Figure \ref{fig:comparison_spectrum} shows the best-fit atmospheric models and the observed spectra for both planets. The retrieved parameters are enlisted in Table \ref{tab:table1}, with posterior distributions shown in Figures \ref{fig:atmo_posterior}, \ref{fig:atmo_posterior_b} for planet c and b, respectively.

We include planet mass as a free parameter for both planets in our retrieval setup. We measure a mass (with 1$\sigma$ confidence interval) of 17$^{+13}_{-6}$~M$_{\oplus}$ for planet c. The posterior distribution shown in (Figure \ref{fig:atmo_posterior}) provides a 3$\sigma$ mass upper limit of $\sim$50~M$_{\oplus}$ for planet c. The measured mass of planet c from the spectrum is consistent with previous RV mass estimates for this planet: 20$\pm$10~M$_\oplus$ from \citet{sikora2023} and $<$50~M$_\oplus$ at 3$\sigma$ level of confidence \citep{finoceti2023}. Our models find sub-solar/solar atmospheric metallicity for this planet: logZ/Z$_{\odot}$=-0.16$^{+1.1}_{-0.9}$. Thus, within 1$\sigma$ confidence interval, the metallicity of planet c is between 0.1-10$\times$ solar. We do not find any constraint on the cloud deck opacity, the equilibrium temperature or C/O ratio for planet c.

For planet b, we measure a mass of 8$^{+4}_{-2}$~M$_{\oplus}$ from our free mass retrievals. The mass posterior distribution for planet b (Figure \ref{fig:atmo_posterior_b}) shows a 3$\sigma$ upper limit at $\sim$20~M$_{\oplus}$. The free mass retrieval posterior distribution from \citet{barat2023}, which used a different atmospheric model (\texttt{PetitRadtrans}) compared to the present study, found a mass (and 1$\sigma$ confidence interval) of 9$^{+4}_{-3}$~M$_{\oplus}$ and 3$\sigma$ upper limit of 23~M$_{\oplus}$. The median value of mass found by \citet{barat2023} is consistent with the present results within 1$\sigma$, and the updated mass upper limit is marginally lower. 

%This difference could be due to different thermal structure assumed for the two models. We assume an isothermal atmosphere, whereas \citet{barat2023} assume a non-isothermal Guillot T-P profile \citep{guillot2010} in their retrievals to explain the low methane abundance. Our retrieved isothermal temperatures for planet b are higher (800-900~K) compared to the isothermal temperature retrieved by \citet{barat2023} (600-700~K), because our models tend to push the isothermal temperature beyond the CO/CH$_4$ transition \citep{fortney_2020}. 

The discrepancy arises because of the non-isothermal T-P profile with high interior temperature and vertical mixing to explain the non-detection of methane for this planet in \citet{barat2023}. However, our models assume isothermal atmospheres, as a result retrievals favor higher isothermal temperature (800-900~K) compared to \citet{barat2023} (600-700~K) to push the carbon chemistry into methane poor regime \citep{fortney_2020}. Therefore we can conclude that although the mass upper limits slightly differ between \citet{barat2023} and this work, they can be considered consistent given the different model assumptions.

%Our isothermal model lacks this flexibility, leading to higher temperatures beyond the CH4/CO transition \citep{fortney_2020} to eliminate methane from the upper atmosphere. As a summary, the mass estimate from free mass retrievals are consistent between \citet{barat2023} and the present study.

Our models retrieve a sub-solar atmospheric metallicity (logZ/Z$_{\odot}$=-2$^{+0.7}_{-0.6}$) for planet b. This retrieved atmospheric metallicity is consistent within 1$\sigma$ to metallicity reported in \citet{barat2023}. We found an upper limit to the cloud opacity for this planet: log${\kappa}_{cld}$=-0.6. We do not find any constraints on the C/O ratio for planet b.

We test the impact of keeping the planet mass as a free parameter in our retrievals. To assess this, we perform retrievals by fixing the mass of both planets. For V1298 Tau c we choose two masses for this test: 10 and 50~M$_{\oplus}$ which are at two ends of the mass posterior distribution (Figure \ref{fig:atmo_posterior}) and are also consistent with the mass upper limit from RV studies \citep{finoceti2023}. For planet b, we choose 10 and 23~M$_\oplus$. \citet{barat2023} has shown that the HST spectrum for V1298 Tau b could be explained with a 10~M$_{\oplus}$ model with a highly sub-solar metallicity and higher cloud opacity, as well as with a relatively clear solar metallicity atmosphere with a planet mass of 23~M$_{\oplus}$. We, therefore, choose these two scenarios for this test.  The posterior distributions for these fixed mass retrievals are shown in Figures \ref{fig:10_23_comparison_corner}, \ref{fig:10_50_comparison_corner}, with retrieved parameters in Table \ref{tab:table1}. The retrieved atmospheric metallicity from these fixed mass retrievals are consistent within 1$\sigma$ with the free mass retrieval case for both planet b and c (Table \ref{tab:table1}). For planet c, we find that increasing the mass also leads to a small increase in the retrieved atmospheric metallicity, but for all three cases (free mass, 10~M$_{\oplus}$ and 50~M$_{\oplus}$) the retrieved metallicity lies in the range 0.1-10$\times$solar for 1$\sigma$ confidence interval. For planet b the metallicity is sub-solar for all mass values, however metallicity increases for higher masses. Considering the 1$\sigma$ confidence interval, for the three cases for this planet (free mass, 10~M$_{\oplus}$ and 23~M$_{\oplus}$), the atmospheric metallicity lies within 0.01-1$\times$ solar metallicity.

From the atmospheric retrievals on V1298 Tau b and c, we conclude that both these planets have masses in the Neptune or sub-Neptune regime. The RV mass for V1298 Tau b from \citet{mascareno_2021} is inconsistent with our measurement, but the planet c mass upper limit from RVs is consistent with our mass measurement for planet c.  The mass measurements of both planet b and c from this work are consistent with follow-up RV studies \citep{finoceti2023,sikora2023}. The mass we measure for planet b is consistent with previous analysis of its HST transmission spectrum \citep{barat2023}.  Furthermore, TTV  analysis of this system (Livingston et al, in prep) also find masses of both these planets in the sub-Neptune mass regime. By testing the impact of including planet mass as a free parameter in our retrievals, we conclude that the atmosphere of V1298 Tau b appears to have metallicity that is lower than or equivalent to solar composition. For V1298 Tau c, the constraints on the atmospheric metallicity are not as precise compared to planet b; however, the posterior distribution for the metallicity peaks at solar composition with 1$\sigma$ confidence interval between 0.1-10$\times$solar. Therefore, it is possible that both these planets are young Neptunes or sub-Neptunes in terms of their mass. But their atmospheric metallicity is much lower (at least 1 order of magnitude) than what is typically found for known mature Neptune or sub-Neptune planets \citep[e.g. see][]{bean2010,desert2011a,kempton2012,morley2013,kreidberg14,libby_robert2020,bean2021}.

\begin{table}
    \centering
    \caption{Derived parameters from atmospheric retrievals run using ATMO (See section \ref{subsec:atmo}) for planet c and planet b.}

%    \centering
    \resizebox{\columnwidth}{!}{\begin{tabular}{c| c | c | c | c}
       Planet & Parameter  & Free mass retrieval (ATMO) & Fixed mass (10~M$_\oplus$) & Fixed mass (50~M$_\oplus$)  \\
       \hline
       \hline
     &  logZ [solar]   & -0.16$_{-0.94}^{1.15}$ & -0.02$_{-1.2}^{1.3}$ & -0.001$_{-1.08}^{1.05}$\\
      
      & R$_p$ [$R_{J}$]   & 0.48$_{-0.01}^{0.01}$ & 0.48$_{-0.01}^{0.03}$  & 0.48$_{-0.01}^{0.01}$ \\
     V1298 Tau c &  T$_{eq}$ [K]  & 947$_{-80}^{+104}$ & 921$_{-79}^{+113}$ & 985$_{-105}^{+95}$ \\
        
        & log${\kappa}_{cld}$ & 0.26$_{-0.83}^{+1.03}$ & 0.73$_{-0.94}^{+0.81}$ & -0.11$_{-0.60}^{+0.97}$ \\
        &  C/O  & 0.06$_{-0.05}^{+0.2}$ & 0.07$_{-0.06}^{+0.26}$ & 0.06$_{-0.04}^{+0.22}$ \\
        & Mass [M$_\oplus$] & 17$_{-6}^{+13}$ & 10 (fixed) & 50 (fixed)  \\
   %    \hline
       \hline
       \hline
       Planet & Parameter  & Free mass retrieval (ATMO) & Fixed mass (10~M$_\oplus$) & Fixed mass (23~M$_\oplus$)  \\
       \hline
       \hline
       &  logZ [solar]   & -2.04$_{-0.59}^{0.69}$ & -1.80$_{-0.7}^{1.07}$ & -0.86$_{-0.85}^{1.05}$\\
      
      & R$_p$ [$R_{\oplus}$]   & 0.84$_{-0.01}^{0.02}$ & 0.84$_{-0.01}^{0.03}$  & 0.84$_{-0.01}^{0.01}$ \\
     V1298 Tau b &  T$_{eq}$ [K]  & 889$_{-180}^{+150}$ & 806$_{-210}^{+196}$ & 824$_{-125}^{+137}$ \\
        
        & log${\kappa}_{cld}$ & 1.03$_{-0.75}^{+0.61}$ & 0.88$_{-0.98}^{+0.68}$ & 0.25$_{-0.62}^{+1.02}$ \\
        &  C/O  & 0.08$_{-0.06}^{+0.18}$ & 0.05$_{-0.03}^{+0.18}$ & 0.04$_{-0.03}^{+0.15}$ \\
        & Mass [M$_\oplus$] & 8$_{-2}^{+4}$ & 10 (fixed) & 23 (fixed)  \\
   %    \hline
       \hline
    \end{tabular}}
%    \caption{Derived parameters from atmospheric retrievals run using ATMO (See section \ref{subsec:atmo}) for planet c and planet b. We ran retrievals keeping both planet mass fixed and as a free parameter in our retrievals. The corresponding posterior distributions are shown in Figure \ref{fig:atmo_posterior}, Figure \ref{fig:atmo_posterior_b}, Figure \ref{fig:10_50_comparison_corner}, Figure \ref{fig:10_23_comparison_corner}. A description of the models used for the retrievals is given in Section \ref{subsec:atmo}. The quoted values are median and 68\% confidence intervals from the retrieved posteriors.}
\tablefoot{We ran retrievals keeping both planet mass fixed and as a free parameter in our retrievals. The corresponding posterior distributions are shown in Figure \ref{fig:atmo_posterior}, Figure \ref{fig:atmo_posterior_b}, Figure \ref{fig:10_50_comparison_corner}, Figure \ref{fig:10_23_comparison_corner}. A description of the models used for the retrievals is given in Section \ref{subsec:atmo}. The quoted values are median and 68\% confidence intervals from the retrieved posteriors.}
    \label{tab:table1}
\end{table}

\subsection{Haze formation in the atmosphere of V1298 Tau b and c} \label{subsec:hazy models}

Hazes have been found to be ubiquitous in exoplanet atmospheres \citep{pont2008,desert2011a,kempton2012,morley2013,kreidberg14,knutson2014,libby_robert2020,kempton2023}. The temperature of V1298 Tau b and c is expected to be between 600-1000K, which has been considered a sweet spot for haze formation \citep{Brande2024}. Furthermore, the host star is young and active, producing copious amounts of UV radiation, which is an important ingredient for haze formation \citep{kawashima2018,kawashima2019,gao2020b,yu2021}.

We simulated the transmission spectrum of V1298 Tau b and c, including hazes. We assumed tholin like composition for the hazes \citep{khare1984}. The haze microphysics models were simulated using the models described in \citet{kawashima2018}, following the approach of \citet{kawashima2019}. First, using a photochemical model the abundance of gaseous species was calculated. The XUV spectrum of V1298 Tau was adopted from \citet{Duvvuri2023}. Haze production rates (or efficiency of haze formation) are uncertain; therefore, we assumed a certain fraction of precursor molecules to be dissociated to form haze monomers. The haze precursor molecules are assumed to be CH$_4$, HCN and C$_2$H$_2$ in our model. We derived vertical density and radius distributions of the haze particles following \citet{kawashima2018}. We assumed solar elemental abundance ratio and other parameters such as eddy diffusion coefficient,
monomer radius, internal density, and refractive index of haze particles similar to \citet{kreidberg2022}.

 We assume 685~K and 979~K as the isothermal temperatures for these models for planet b and c, respectively (Table \ref{tab:b_c_comparison}). Since the haze models are computationally expensive, we create models for two masses of planet b and c each. For planet b, we choose two masses: 10~M$_{\oplus}$ (consistent within 1$\sigma$ to the median of the ATMO retrieval mass posterior distribution for planet b) and 23~M$_{\oplus}$ (3$\sigma$ upper limit from \citet{barat2023}). For planet c, we choose 10~M$_{\oplus}$, and 20~M$_{\oplus}$ (consistent within 1$\sigma$ to the median of the ATMO retrieval mass posterior distribution for planet c). We choose 10~M$_{\oplus}$ for planet c because from ongoing TTV studies, it is likely that the mass of planet c is in the sub-Neptune regime (Livingston et. al in prep). 

In Figure \ref{fig:comparison_spectrum} we show the simulated transmission spectra assuming a 1$\%$ haze production efficiency for mass of 8 and 17~M$_{\oplus}$ for planet b and c respectively for the ease of comparison with the ATMO models. We re-scale the planet c and b models computed with masses of 20 and 10~M$_{\oplus}$ respectively to 17 and 8M${_\oplus}$ models, assuming that a small change in the planet mass will not significantly affect the haze models. The atmospheric signal (which is a relative measurement) is inversely proportional to the planet mass \citep{heng2017}, through the scale height. Assuming this principle, we re-scale the haze grid models to different mass values.   While significant haze formation can be ruled out for V1298 Tau b, high haze formation rates could explain the observed spectrum of V1298 Tau c.

 The haze formation efficiency is an uncertain parameter in our models. Furthermore, the mass of the planets is also uncertain. So, to test the effects of this parameter on the hazy atmospheric models, we create a grid of models at different haze formation efficiency and planet mass. In Figure \ref{fig:data_haze_model_comparisons} we show the spectra of both planets with different haze formation efficiencies: 10$^{-4}$ and 10$^{-5}$ for planet b and 10$^{-5}$ and 10$^{-2}$ for planet c. For V1298 Tau b, irrespective of the mass, haze formation efficiencies as low as 10$^{-4}$ can be ruled out as these do not reproduce the large water absorption feature. At 8~M$_{\oplus}$, a haze formation efficiency of 10$^{-5}$ can potentially explain the water absorption feature for planet b, but introduces a strong slope in the continuum which is not seen in the observed spectrum. For V1298 Tau c, the haze formation rates cannot be constrained as haze formation efficiencies as low as 10$^{-5}$ can explain the spectrum. However, for the 10~M$_{\oplus}$ for planet c we can rule out low haze formation efficiencies ($<$10$^{-5}$) at 3.5$\sigma$ level of confidence. Therefore, we conclude that V1298 Tau b has a haze formation efficiency lower than 10$^{-5}$, whereas for V1298 Tau c the atmosphere is likely to have higher haze formation than its sibling planet.

\subsection{Relative transmission spectroscopy between planet b and c} \label{relative transmission spectrum}
In Figure \ref{fig:comparison_spectrum} (lower panel), we present the first relative transmission spectrum between two planets within the same system. By relative transmission spectrum we mean the ratio of transit depths between two planets as a function of wavelength. In principle, the relative transmission spectrum is independent of the stellar radii and limb darkening effects, assuming that both planets share similar limb darkening, i.e they transit similar parts of the stellar disk. V1298 Tau b and c have impact parameters consistent with each other within 1$\sigma$ \citep{david2019b} and relatively low mutual inclination \citep{marshall2021}.  Stellar inhomogeneities are known to contaminate transmission spectra \citep{desert2011a,rackham2019}, which has also been highlighted in recent JWST observations\citep{may2023,moran2023,lim2023}. A relative transmission spectrum between multiple planets in the same system, theoretically is independent of stellar contamination, assuming similar stellar contamination across both transits. This enables a better relative comparison of abundances between the planets without precise information on the stellar properties, including metallicity, elemental abundance ratios, and limb darkening. However, in the case of V1298 Tau, an active star, changes in the stellar surface between observation epochs may still introduce contamination to the relative spectrum extracted at different times.

\section{Discussion} \label{sec:discusison}

\subsection{Relative orbital distances and stellar density} \label{stellar density}

In multi-planet systems, it is possible to fix one planet's semi-major axis based on another planet's measured semi-major axis, since they are related to their orbital period ratio. However, due to known TTVs in this system, we refrain from fixing the semi-major axis in this analysis. The semi-major axis ratio from independent fits to the white light curves of V1298 Tau b and c is 2.08$\pm$0.1 and agrees with the expected value from their known orbital periods. Assuming circular orbits, the semi-major axis can be used to estimate stellar density \citep{Sandford2017}. From the fits to our white light curves, we estimate the stellar density from the planet b and planet c transit light curve to be 0.70$\pm$0.01~gm/cm$^3$ and 0.67$\pm$0.1~gm/cm$^3$. These values align with the published stellar density for V1298 Tau (0.68$\pm$0.1~gm/cm$^3$, \citet{david2019}). The relatively low density of V1298 Tau compared to similar mass main-sequence stars indicates that it is inflated and currently undergoing an evolutionary phase, and is likely to end up as a sun-like star as it approaches the main-sequence \citep{david2019b}.

\subsection{Limb darkening of V1298 Tau} \label{v1298_limb_darkening}

Modelling stellar limb darkening is crucial for deriving transmission spectra. Prior studies often used fixed limb darkening for main-sequence stars based on theoretical stellar structure models \citep[e.g, see][]{sing2016}. \citet{barat2023} fitted a linear limb darkening coefficient for V1298 Tau, revealing significantly higher limb darkening compared to theoretical models for a main-sequence star with a similar spectral type (K1). The limb darkening coefficients have been derived from fitting the transit light curves of V1298 Tau b observed using HST/WFC3 G141. These observations had both ingress and egress coverage and allowed precise constraints on the limb darkening model. However, for simplicity we choose a linear limb darkening model, rather than other limb darkening models with more free parameters, since HST does not provide continuous time series. Higher limb darkening of pre-main sequence stars compared to theoretical models for mature stars has previously been noted for $\beta$-Pictoris \citep{landman2024}. The fitted limb darkening coefficients and a theoretical linear limb darkening model is shown in Figure \ref{fig:limb_darkening_comparison}. This is the first direct measurement of limb darkening on a weak-lined T-Tauri star.

In Section \ref{stellar density}, we highlighted that V1298 Tau's density is lower than that of similar-mass main-sequence stars. Furthermore, it has a relatively high magnetic field strength of the order of 100-300~Gauss \citep{finoceti2023}. High magnetic fields have been proposed as a potential cause for discrepancies between observed and theoretical limb darkening models \citep{kostogryz2024}. Therefore, a combination of structural differences and higher magnetic fields compared to main-sequence stars, could explain the differences we find with the main-sequence limb darkening model.

While leveraging light curves from multiple planets in the same system could potentially reduce limb darkening coefficient uncertainties, the absence of ingress or egress coverage in the light curves of planet c led us to fix the limb darkening coefficients to the values from \citet{barat2023}. A comparison of the transmission spectrum of planet c derived using fixed and free limb darkening is shown in Figure \ref{fig:fix-free-comparison}. For the both cases, we assume a linear limb darkening law. For the free limb darkening case, we put uniform priors between 0 and 1 on the linear limb darkening coefficient. The transit depth uncertainties are higher by 5-10\% for each spectroscopic bin in the latter case. Residuals in both cases are approximately 1.2 times the expected photon noise, indicating that stellar variability and instrumental systematics from the first orbit are likely to be the dominant sources of noise in our case. However, the median subtracted transit depths are consistent within 1$\sigma$ for all spectroscopic bins (Fig \ref{fig:fix-free-comparison}). For the atmospheric retrievals we use the transmission spectrum derived using the fixed limb darkening case. In Figure \ref{fig:fix-free-comparison} we also show a transmission spectrum for planet c derived  by fixing the linear limb darkening coefficient to the EXOCTK model shown in Figure \ref{fig:limb_darkening_comparison}. We find that the median subtracted transmission spectra for this scenario is consistent within 1$\sigma$ for all spectroscopic bins with the fixed limb darkening case used for the analysis in this paper.

\subsection{Comparative exoplanetology between V1298 Tau b and c}

In Figure \ref{fig:comparison_spectrum} we show a comparison between the atmospheres of V1298 Tau b and c. Comparing the spectra in terms of the scale height of both planets (Figure \ref{fig:comparison_spectrum}, upper panel), shows that these planets could host atmospheres of similar nature. To compute the scale height we used the best fit mass from the ATMO posteriors for both planets (Table \ref{tab:table1}). V1298 Tau c has a smaller scale height ($\sim$ by a factor of 4) compared to V1298 Tau b. Atmospheric  retrievals show that the masses of both these planets are likely to be in the Neptune/sub-Neptune regime (Section \ref{subsec:atmo}). These models also show that the atmospheric metallicities of both these planets are sub-solar to solar. Mature Neptune/sub-Neptune mass planets, are known to have metal-rich/hazy (>$100\times$solar) atmospheres \citep{bean2010,desert2011a,kempton2023,madhusudhan2023,roberts2020,bean2021}. Thus, V1298 Tau b and c, which are the two young Neptune/sub-Neptune progenitors, appear to have distinct atmospheres compared to their mature counterparts.

Efficient haze formation ($>$10$^-5$) can be ruled out for V1298 Tau b (Figure \ref{fig:data_haze_model_comparisons}), given the large atmospheric absorption feature. However for planet c atmospheric hazes, can explain the observed spectrum. In the scenario that the mass of planet c is around 17~M$_{\oplus}$, we cannot constrain the haze formation efficiency for planet c, but for lower masses ($\sim$10~M$_{\oplus}$), very low haze formation efficiency ($<$10$^{-5}$) can be ruled out. Therefore, planet c could have a higher haze formation efficiency compared to its sibling planet. Haze formation has been linked to the UV flux of the host star; a higher UV flux results in higher monomer production rates for organic hazes \citep{kawashima2019}. Higher atmospheric metallicity could also lead to higher haze formation. The higher haze formation in planet c compared to its sibling planet could be due to difference in metallicity, temperature and incident UV flux of the two planets.

\subsection{Comparing the early evolution of V1298 Tau b and c} \label{subsec:evolution comparison}

We ran evolutionary models for V1298 Tau b and c. These models start from the time of disk dispersal, and model the structural and thermal evolution, including atmospheric mass loss due to photoevaporation \citep{owen2013}. We terminate atmospheric mass loss at 100~Myr. We run models assuming both a core-envelope interior structure, as well as with a gradually mixed interior. We have used the evolutionary models presented in \citet{vazan2022}. A brief description of the models is provided in Appendix \ref{appendix: evolution}. Figure \ref{fig:evolutionary models} shows a possible mass and radius evolution tracks for these planets.

We set up the initial conditions of our models such that these models can match the mass and radius of the planets at the current age (23~Myr). For planet c we assume 5.24R$_{\oplus}$ and 17~M$_{\oplus}$ (median value of mass from ATMO retrieval). For planet b we assume 9.95~R$_{\oplus}$ and show two models: a 10~M$_{\oplus}$ model which is consistent within 1$\sigma$ to the median value from our ATMO retrievals for planet b, and a 23~M$_{\oplus}$ model which is the 3$\sigma$ mass upper limit from \citet{barat2023}. We assume that the 10~M$_{\oplus}$ model should be similar to a 8~M$_{\oplus}$ model. For this calculation we do not account for the uncertainty in the mass of the planets, therefore, the models should be considered as potential evolutionary tracks.

Based on these models, V1298 Tau c could have started out with $\sim$20~M$_{\oplus}$ with a 25\% by mass H/He envelope. Currently, it hosts $\sim$10\% by mass H/He in its envelope and after 100~Myr of evolution it ends up as a Neptune mass planet with $<$1\% by mass H/He envelope. After 100~Myr this planet enters into a phase of gradual thermal contraction and ends up with a radius between 2.5-3R$_{\oplus}$ (typical for Neptunes/sub-Neptunes). However, for V1298 Tau b, atmospheric mass loss is not as dominant, because it is further away from the host star, thereby, receiving $\sim$25\% of the stellar flux compared to planet c. It starts off with $\sim$8.5~M$_{\oplus}$ and 23.5~M$_{\oplus}$ for the two models presented here, and loses less than 1~M$_{\oplus}$ in the course of its evolution. However, it is important to note here that \citet{barat2023} report a high internal temperature ($\sim$400K) for V1298 Tau b from disequilibrium chemistry models of its atmosphere, which has not been considered in our evolution models. Higher than expected internal temperature for planet b could lead to a puffed up atmosphere and enhanced atmospheric mass loss compared to our models. 

The mass loss rates are strongly dependent on the current masses of the planets. In the scenario that planet b and/or c has mass lower than the estimates used for these calculations the H/He mass fraction in their envelope would be higher, leading to higher mass loss rates. In extreme scenarios, these planets could be completely stripped of their primordial atmospheres. 

While the final mass and radius depend on photoevaporation efficiency and the star's activity level, from these models, we can conclude that due to its proximity to the star, V1298 Tau c is likely to lose a larger fraction of its H/He envelope compared to planet b. Therefore, even if both planets have similar masses now, their early evolution could lead them to different mass-radius regimes as well as on different ends of the radius valley.

Our models assume solar metallicity for the envelope opacity. However, the transmission spectrum of V1298 Tau c could also be explained using atmospheric hazes, unlike V1298 Tau b (Figure \ref{fig:comparison_spectrum}. Young planets are expected to cool and contract over time \citep{kubyshkina2020,linder2019}. The presence of high altitude hazes can increase the atmospheric opacity, therefore increasing the cooling and contraction timescale \citep{lee2015}, leading to higher photoevaporative mass loss. Thus, a hazy atmosphere of V1298 Tau c as opposed to a clear atmosphere of V1298 Tau b, could also lead to slower cooling and higher mass loss rates for planet c.

\subsection{Comparing the atmospheres of V1298 Tau b and c with the exoplanet population}

\begin{figure}
    \centering
    \includegraphics[width=\linewidth]{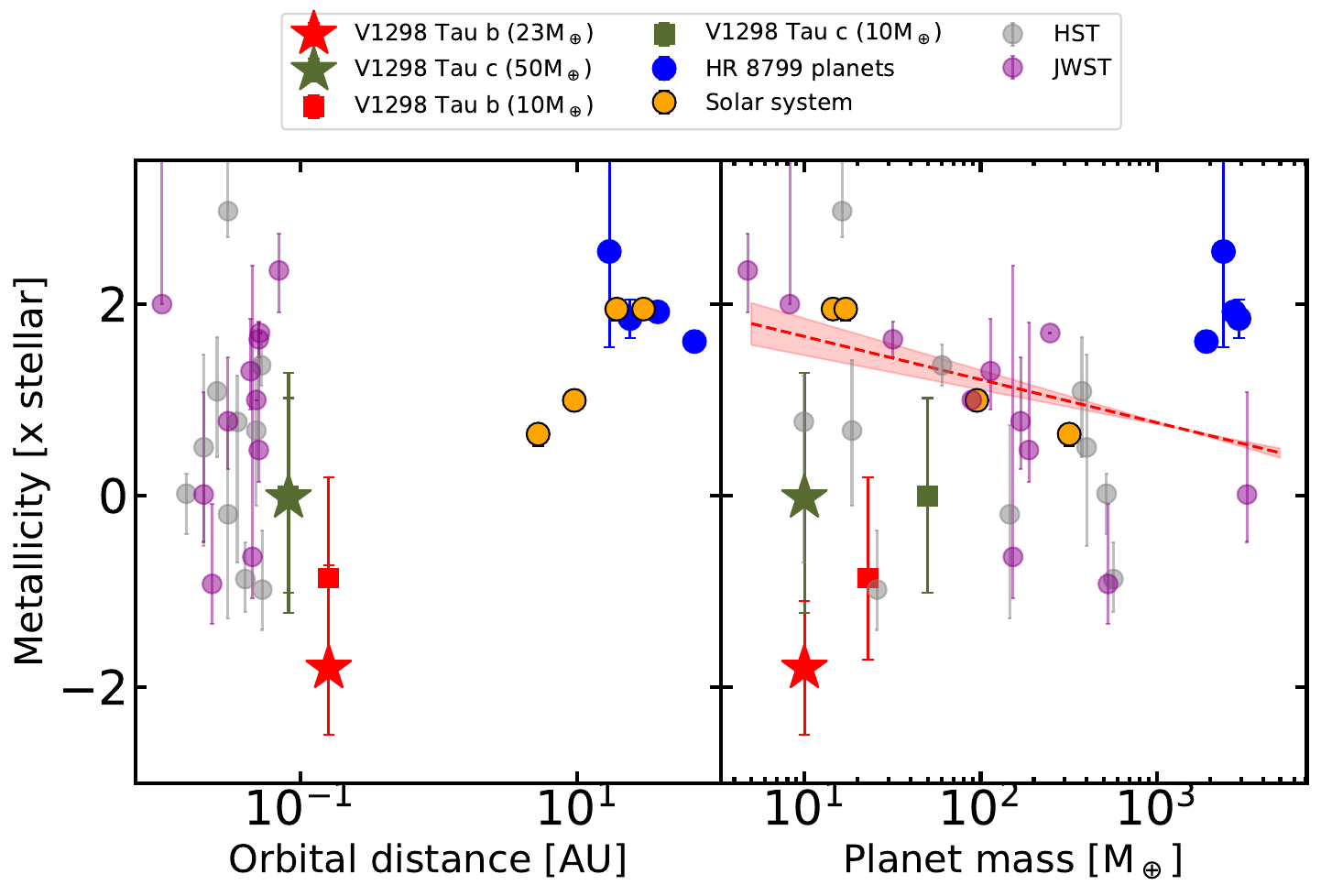}
    \caption{Comparison of the retrieved atmospheric metallicity of V1298 Tau b and c with those of mature exoplanets, with the solar system planets and with young directly imaged planets orbiting HR 8799 (30~Myr old). The left panel shows these planets in the metallicity-orbital distance plane and in the right panel the mass-metallicity distribution of these planets. The atmospheric metallicity of the two V1298 Tau b and c planets have been derived from the ATMO retrievals (Table \ref{tab:table1}). The stars show metallicities for retrievals performed using 10~M$_\oplus$ for both planets and the squares for retrievals using 23 and 50~M$_\oplus$ for planet b and c respectively.  The metallicities of the HR 8799 system (blue circles) have been obtained from \citet{nasedkin2024}. The grey points show metallicity constraints from pre-JWST observations (mostly driven by HST) taken from \citet{wakeford_dalba} and the magenta points show metallicities derived from JWST observations. Table \ref{tab:jwst_metallicity_table} shows the data for the planets shown in magenta with corresponding references. For planets which appear both in the sample of \citet{wakeford_dalba} and Table \ref{tab:jwst_metallicity_table}, we plot the latter. For planets where metallicity is estimated from grids, we do not plot errorbars. A comparison between the measured metallicties between the V1298 Tau and HR8799 systems show that these planets have very different atmospheric metal enrichment, indicative of differences in their formation mechanism, disk properties, and location. The red dashed line in the right panel shows a mass-metallicity relation derived in \citet{thorngren_2019}. The V1298 Tau planets have  metal-poor atmospheres by an order of magnitude compared to their mature counterparts and observed mass-metallicity trend among exoplanets.
   }
    \label{fig:metallicity-distance}
\end{figure}

Figure \ref{fig:metallicity-distance} shows V1298 Tau b and c on the mass-metallicity and metallicity-orbital distance planes. The observed low atmospheric metallicity (solar/sub-solar) for these planets, deviates from the mass-metallicity trend predicted by core-accretion models \citep{boddenheimer1986,pollack1996} and reported for exoplanets \citep{welbanks_19,thorngren_2019}. The atmospheric metallicities of planet b and c appears to be at least one order of magnitude lower compared to mature Neptune/sub-Neptunes. This could potentially be explained by ongoing evolutionary processes that transform their current metal-poor atmospheres of V1298 Tau b and c into the metal-rich atmospheres, as found in known mature Neptune/sub-Neptune planets \citep{desert2011a,kempton2012,morley2013,libby_robert2020,bean2021,kempton2023,madhusudhan2023}. One such possible mechanism has been suggested, where Neptune/sub-Neptunes are naturally born with metallicity gradients in their envelope \citep{ormel2021}. Atmospheric mass loss during the early evolutionary phase removes the upper layers of the atmosphere, revealing the inner metal-rich layers \citep{fortney2013,vazan2022}. Thus, this mechanism could reconcile the the two Neptune/sub-Neptune progenitors, V1298 Tau b and c with their mature counterparts.

We compare the V1298 Tau system with the $\sim$30~Myr old HR 8799 system \citep{marois2010} with four directly imaged planets (HR8799 b, c , d and e) with masses 7$^{+4}_{-2}$, 10$^{+3}_{-3}$, 10$^{+3}_{-3}$, 10$^{+7}_{-4}$~M$_{J}$, respectively \citep{marois2010,lacour2019}. \citet{nasedkin2024} using ground based spectra from the VLTI/GRAVITY program,  report super-stellar($\sim$100 times stellar) atmospheric metallicity for these planets. Assuming a -0.65~dex metallicity for HR8799 \citep{swastik2021} we convert the retrieved atmospheric metallicity to stellar unit.

The HR 8799 planets exhibit higher atmospheric metallicity compared to the mass-metallicity trend \citep{thorngren_2019} predicted by core-accretion formation, while V1298 Tau planets fall below this trend. HR 8799 is a system of widely separated massive gas giants, whereas V1298 Tau likely consists of compact, low-mass Neptune/sub-Neptune progenitors. Both systems likely host primordial atmospheres. The different mass regimes suggest HR 8799 planets underwent runaway gas accretion \citep{pollack1996} or gravitational collapse \citep{boss1997}, unlike V1298 Tau planets. Formation theories of Neptune/sub-Neptunes suggest that these planets are born in situ within the water ice-line in dusty or depleting disks to prevent runaway gas accretion \citep{lee2014,lee_chiang2016}. Wide orbit gas giants likely formed beyond the ice lines, allowing for runaway accretion \citep{pollack1996} or gravitational instability \citep{boss1997,Dodson-Robinson2009}, but these mechanisms alone do not explain HR 8799's high metallicity, necessitating the accretion of volatile-rich solids \citep{nasedkin2024}. Formation location relative to ice lines could lead to differences in volatile enrichment, as beyond the ice-line, the disk contains volatile-rich solids, enhancing the metal content of primordial atmospheres \citep{oberg_2011,bitsch2019}.

\subsection{Formation and evolution of multi-planet systems}

The mass of V1298 Tau b and c indicate that they are likely Neptune/sub-Neptune progenitors, which are in a near mean-motion resonance (1:2:3). Atmospheric retrievals suggest both these planets could have relatively low metallicity atmospheres. Population synthesis studies \citep{benz2014,mordasini2018} show that low mass planets, often form in multi-planet systems in mean motion resonance \citep{lambrechts2019}. \citet{emsenhueber2023} classified systems with multiple low-mass planets into two types: Class I: inner dry/rocky planets formed in situ with outer icy sub-Neptunes migrating inwards, and Class II: multiple sub-Neptunes forming beyond the water ice line and migrating inwards. In the first scenario, inner planets have metal-poor atmospheres, whereas, in the second, all planets are metal-rich.

The relatively low atmospheric metallicity of V1298 Tau b and c suggests a formation similar to class I planets . However, these young planets' atmospheric metallicity may evolve. In the pebble accretion scenario, a metallicity gradient forms in the protoplanet’s envelope \citep{boddenheimer2018,ormel2021}, with metal-rich layers hidden deep inside. Over time, atmospheric mass loss can strip volatile-poor upper layers, revealing deeper metal-rich atmospheric layers\citep{fortney2013,vazan2022}, aligning them with mature sub-Neptune/super-Earth population. In this case, the V1298 Tau system could also have a formation similar to the class II systems.

\begin{figure}
    \centering
    \includegraphics[width=\linewidth]{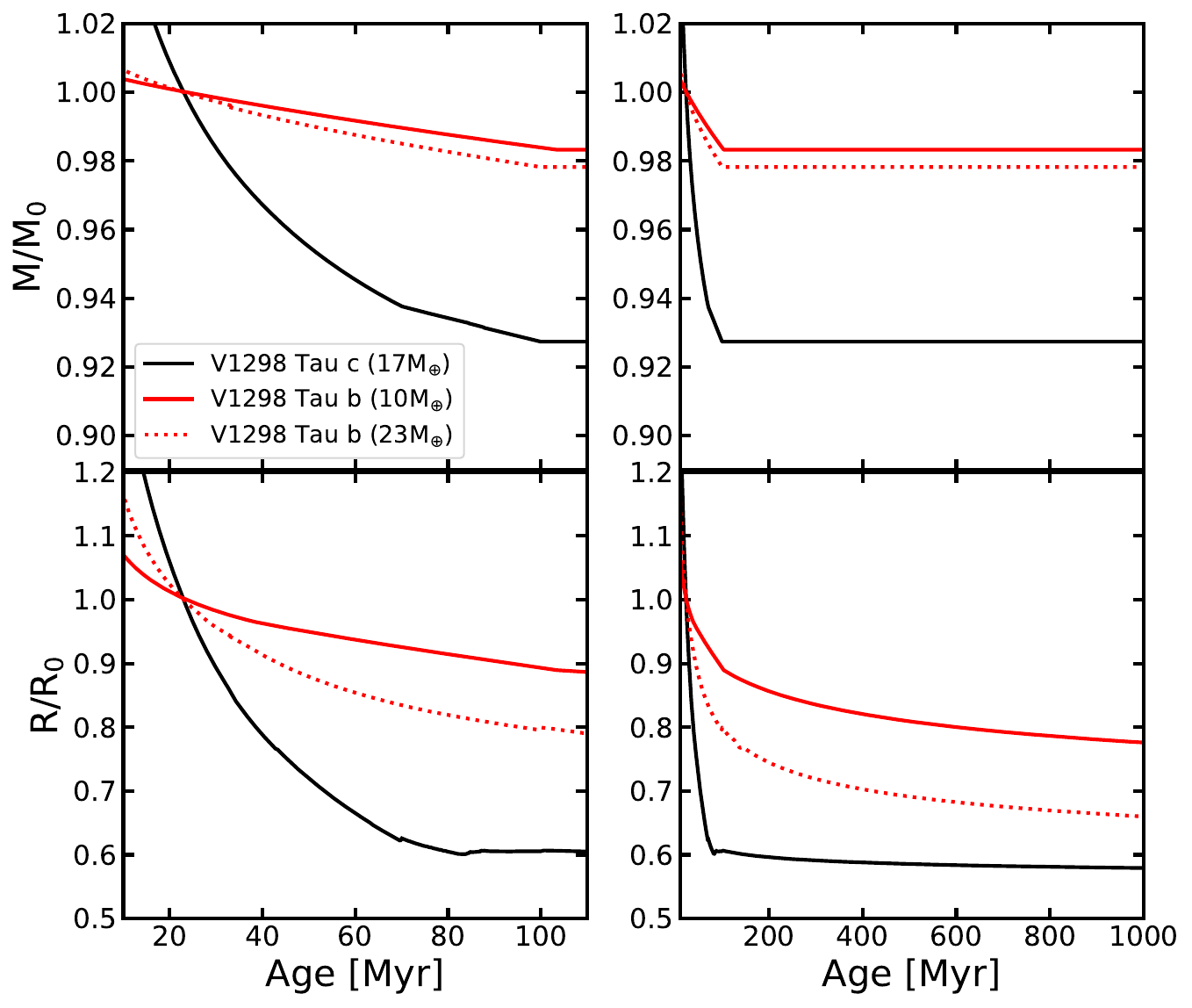}
    \caption{Potential mass (upper panel) and radius (lower panel) evolutionary tracks for V1298 Tau c (black) and V1298 Tau b (red). M$_0$ and R$_0$ refers to the current mass and radius respectively. For planet c we assume 5.24R$_{\oplus}$ and 17~M$_{\oplus}$ (median value from ATMO atmospheric retrieval) as the current radius and mass. For planet b we assume 9.95~R$_{\oplus}$ as the current radius. We show models for two values for the current mass for planet b: 10~M$_{\oplus}$ (red solid line) which is consistent within 1$\sigma$ to the mass measured from our ATMO retrievals (8$^{+4}_{-2}$ M~$_{\oplus}$) and 23~M$_{\oplus}$ which is the 3$\sigma$ upper limit from \citet{barat2023}. The left panels zoom in on the first 100~Myr and the right panels show the evolutionary tracks for 1~Gyr. The evolutionary models described in Appendix \ref{appendix: evolution} calculate the mass and radius evolution due to cooling and atmospheric mass loss. We used evolutionary models described in \citet{vazan2022}. For further discussion see Section \ref{subsec:evolution comparison}. These evolutionary models represent one possible scenario for the early evolution of both these planets and do not take into account the planet mass uncertainty.}
     
    \label{fig:evolutionary models}
\end{figure}

If V1298 Tau c has a hazy atmosphere, it might be more metal-rich than the outer planet b, which doesn't fit either class I or class II systems. Class I systems have inner volatile-poor planets, while class II systems have planets with similar compositions. Due to its proximity to the star, V1298 Tau c is likely to lose more of its primordial envelope compared to V1298 Tau b  \ref{fig:evolutionary models}. Thus, V1298 Tau b and c might have formed like class II systems, with volatiles hidden deep inside, but atmospheric mass loss caused planet c's metallicity to evolve faster than planet b.

\begin{figure}
   \centering
    \includegraphics[width=\linewidth]{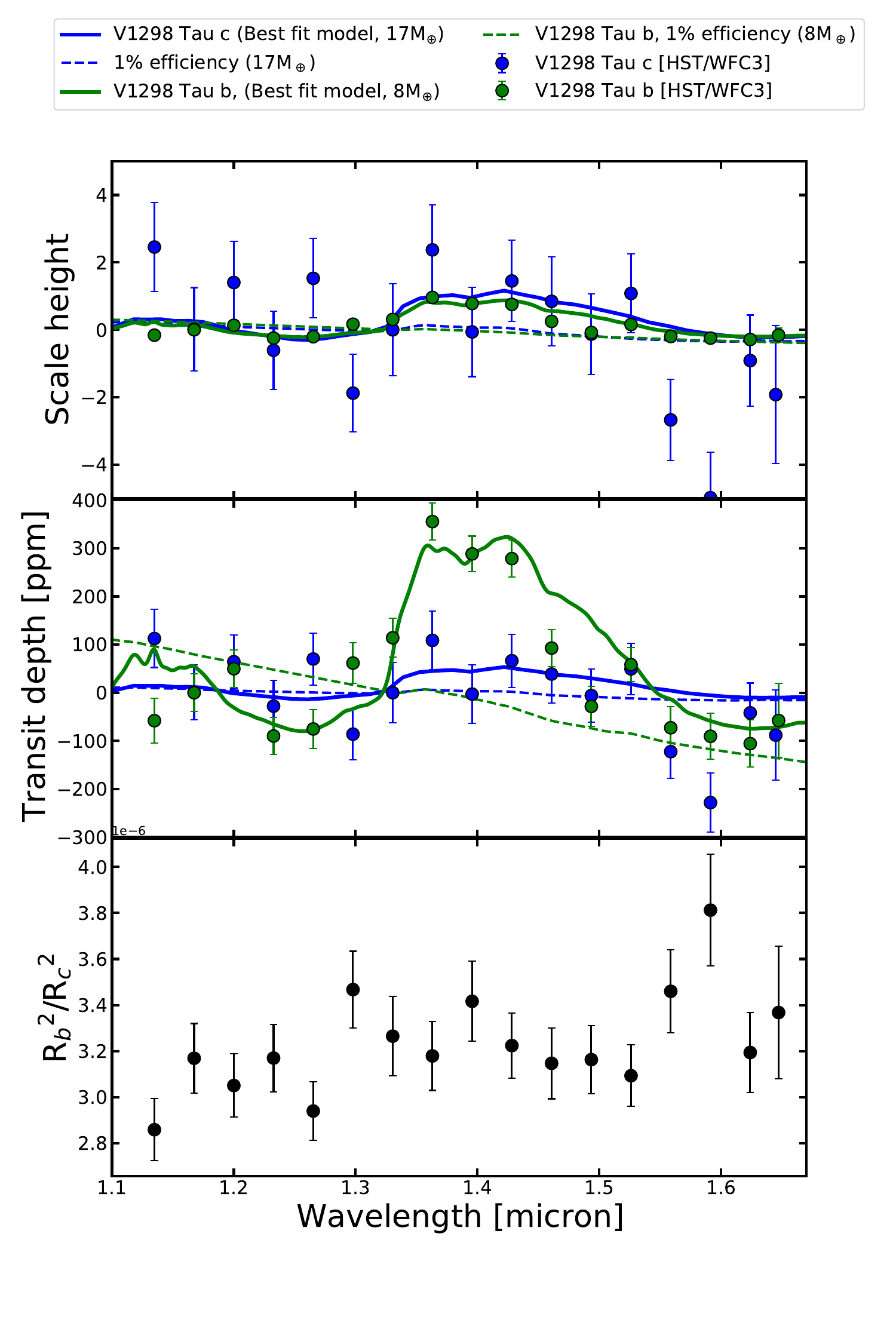}
    \caption{ Transmission spectrum of V1298 Tau c (blue points) shown in comparison with its sibling planet V1298 Tau b (green points). The spectra have been derived by subtracting the median of the observed spectrum. The upper panel shows the spectra in terms of atmospheric scale height and the middle panel in terms of transit depth. We assumed T$_{eq}$=979~K and 17~M$_{\oplus}$ for planet c and 685~K and 8~M$_\oplus$ for planet b to calculate the scale height for each planet. We assumed H/He dominated primordial atmospheres and assumed a mean molecular weight of 2.33. The green and blue solid lines show the best fit ATMO models from our retrievals for planet b and c with 8 and 17~M$_{\oplus}$ respectively (Section \ref{subsec:atmo}). The dashed lines show hazy atmospheric models for both planets simulated using the formalism presented in \citet{kawashima2018} (See Section \ref{subsec:hazy models}). We show hazy atmospheric models with the same mass as the best fit models and have been generated by re-scaling the hazy atmospheric model grids described in Section \ref{subsec:hazy models}. The lower panel shows the relative transmission spectrum (i.e the ratio between transmission spectra) between V1298 Tau b and V1298 Tau c (See Section \ref{relative transmission spectrum}).   }
    \label{fig:comparison_spectrum}
\end{figure}

\begin{figure}
    \centering
    \includegraphics[width=\linewidth]{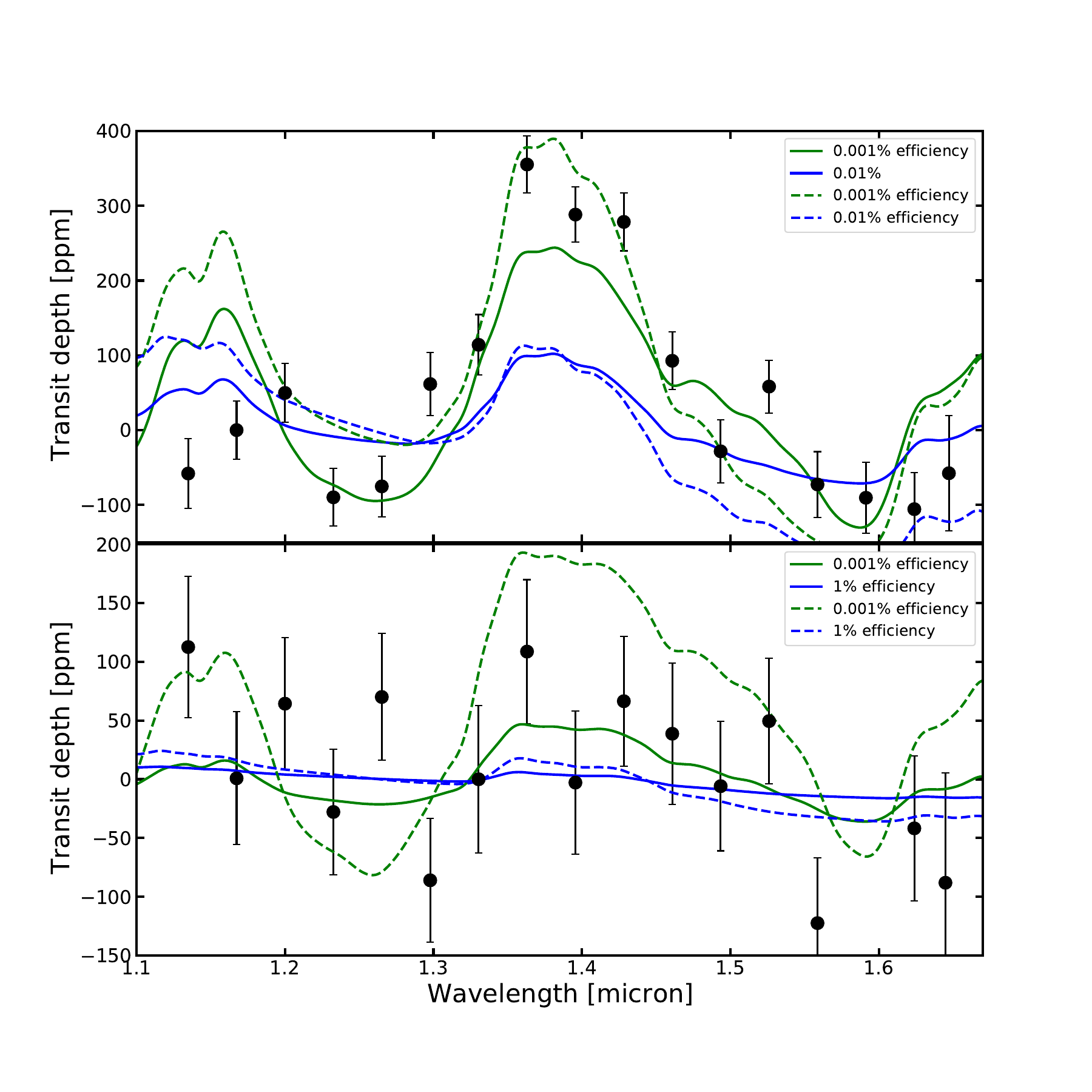}
    \caption{Comparison between observed HST transmission spectrum of V1298 Tau b (upper panel) and V1298 Tau c(lower panel) with haze grid models of different haze formation efficiency and planet mass. Upper panel: Solid lines represent models assuming 23~M$_{\oplus}$. Dashed lines represent models with 8~M$_{\oplus}$, which have been re-scaled from 10~M$_{\oplus}$ haze grid model of planet b. Blue and green represent haze formation efficiency of 0.01\% and 0.001\% respectively. Lower panel: Solid lines represent models assuming 17~M$_{\oplus}$ re-scaled from 20~M$_{\oplus}$ haze grid model for planet c. Dashed lines represent models with 10~M$_{\oplus}$. Blue and green represent haze formation efficiency of 1\% and 0.001\% respectively. }
    \label{fig:data_haze_model_comparisons}
\end{figure}

\section{Conclusion} \label{sec:conclusion}

In this paper, we perform the first comparative study of atmospheres of multiple planets in the same transiting multi-planet system. We compare the transmission spectra of two exoplanets in the young (20-30~Myr old) V1298 Tau system: V1298 Tau b and c. The transmission spectrum of V1298 Tau c shows an absorption feature that we interpret as due to water vapour (2.5$\sigma$), whereas, the spectra of its sibling, V1298 Tau b (adopted from \citet{barat2023}), shows a clear atmosphere with a large scale height and a water absorption feature at 5$\sigma$ level of significance (Figure \ref{fig:comparison_spectrum}). We measure the masses of both planets from the transmission spectra ($8_{-2}^{+4}$~M$_{\oplus}$ for planet b and $17_{-6}^{+13}$~M$_{\oplus}$ for planet c), together with the chemical abundances using a Bayesian chemical equilibrium retrieval framework using the ATMO atmospheric model. The retrieved mass for planet c is consistent with previous RV estimates \citep{mascareno_2021,sikora2023,finoceti2023}. Our mass determination from the atmospheric scale heights also agree with ongoing TTV studies (Livingston et al, in prep). Our atmospheric models find sub-solar/solar atmospheric metallicity (logZ/Z$_\odot$=-2.04$_{-0.59}^{0.69}$ for planet b and logZ/Z$_\odot$= -0.16$_{-0.94}^{1.15}$ for planet c) for both these planets (Table \ref{tab:table1}), from free planet masses as well as fixed masses in the atmospheric retrievals. Compared to known mature Neptune/sub-Neptunes, which have been found to have metal-rich ($>$100$\times$solar) and/or hazy atmospheres, the atmospheres of the two young Neptune/sub-Neptune progenitors (planet b and c), can be considered metal-poor by at least one order of magnitude. This hints at possible ongoing early evolutionary mechanisms which are expected to enhance the atmospheric metallicity of these planets, such that they are reconciled with their mature counterparts after evolution. Alternatively, the observed spectrum of V1298 Tau c could also be explained by atmospheric hazes. The large observed scale height of V1298 Tau b rules out high haze formation efficiency ($>10^{-5}$, Figure \ref{fig:data_haze_model_comparisons}) for this planet. Higher haze formation in planet c compared to planet b could be due to differences in bulk composition, temperature-pressure profile, and incoming stellar UV flux, which are known to be important factors affecting haze formation in H/He rich atmospheres \citep{kawashima2018,gao2020}. In this scenario, the higher envelope opacity of planet c due to hazes could delay its cooling rate and compared to planet b, which has a clearer atmosphere. Evolutionary models suggest that planet c could lose a large fraction of its current H/He envelope ($\sim$10\% by mass) and end up with $<1$\% by the end of it's evolutionary phase. In contrast, our models applied to planet b, which receives 25\% of the stellar flux received by planet c, do not predict significant mass loss and could end up retaining a substantial H/He envelope after 100~Myr of evolution.

\begin{acknowledgement}

J.M.D acknowledges support from the Amsterdam Academic Alliance (AAA)
Program, and the European Research Council (ERC) European Union’s Horizon 2020 research and innovation program (grant agreement no. 679633;
Exo-Atmos). This work is part of the research program VIDI New Frontiers in
Exoplanetary Climatology with project number 614.001.601, which is (partly)
financed by the Dutch Research Council (NWO).
Y.K. acknowledges the support from JSPS KAKENHI Grant Numbers 21K13984, 22H05150, and 23H01224. We acknowledge publicly available open source softwares which have been used for this work: \texttt{numpy, scipy, matplotlib, emcee, lmfit, pandas, ATMO, PetitRadtrans}
    
\end{acknowledgement}

\

 \newpage
\bibliography{references}

 \appendix

 \section{Description of raw data reduction} \label{appendix:pipleine}

We use the HST/WFC3 data reduction pipeline which has been presented in previous studies \citep{Arcangeli2018a,arcangeli2019,arcangeli2021,jacobs2022, barat2023}. We do not change anything in the pipeline for the analysis of this paper.
 We start our analysis from the `ima.fits' files downloaded from MAST archive. During each exposure, the detector is read multiple times, without flushing out the accumulated charge (non-destructive reads). We create sub-exposures by subtracting consecutive non-destructive reads for each exposure which are reduced separately. This technique has been widely applied in past observations with HST. We apply a wavelength dependent flat-field correction and flag bad pixels with data quality DQ=4, 32, or 512 by \texttt{calwf3}. We apply a local median filter to identify cosmic rays and clip pixels which are more than five median deviations from the median. To account for the dispersion direction drift of the spectrum \citep{wakeford2016} we use the first exposure of a visit as a template and shift the spectrum for an exposure along the dispersion direction to match the template. The maximum shift that we require is 0.1 pixels. We apply optimal extraction algorithm \citep{horne1986} on each subexposure to maximize signal-to-noise ratio. 

 \section{Light curve fitting} \label{light curve fitting}

 HST WFC3 light curves often exhibit strong systematics, such as `hook-like' ramps and visit-long slopes \citep{berta2012,deming13,ranjan2014,tsiaras2016,wakeford2016}. Common practice involves discarding the first orbit due to significantly larger systematics (See Figure \ref{fig:whitelc}) compared to the rest of the visit \citep[e.g. see,][]{wakeford2013,Arcangeli2018a,jacobs2022,barat2023} from the light curve fitting. For the visit of planet b, the first orbit had been discarded \citep{barat2023}. However, for the visit of planet c, we could not remove the first orbit of the visit from the analysis because the transit occurred two hours earlier than expected from the linear ephemeris \citep{Feinstein2022} due to large TTVs and we could only secure one orbit prior to ingress.

 The first-orbit systematics are strongly wavelength dependent \citep{Zhou2020,zhou2017}. Therefore, the spectroscopic light curves cannot be de-trended using a common-mode approach, which had been previously applied to similar observations of V1298 tau b \citep{barat2023}. Therefore, to handle these systematics, we used the physically motivated RECTE charge-trapping model \citep{zhou2017}, which fits the first orbit ramp better than other known methods \citep[e.g,][]{berta2012}. We fixed the mid-transit time using simultaneous TESS observations \citep{Feinstein2022}. V1298 Tau shows strong rotational variability, with an average amplitude of 2$\%$ and a rotational period of 2.83 days \citep{david2019b}. We modeled it with a second-order polynomial for the out-of-transit baseline. See Appendix \ref{subsection:stellar baseline} for more discussion on the baseline modeling. The planetary transit signal was modeled using \texttt{batman} \citep{batman}.

\begin{figure}
    \centering
    \includegraphics[width=\linewidth]{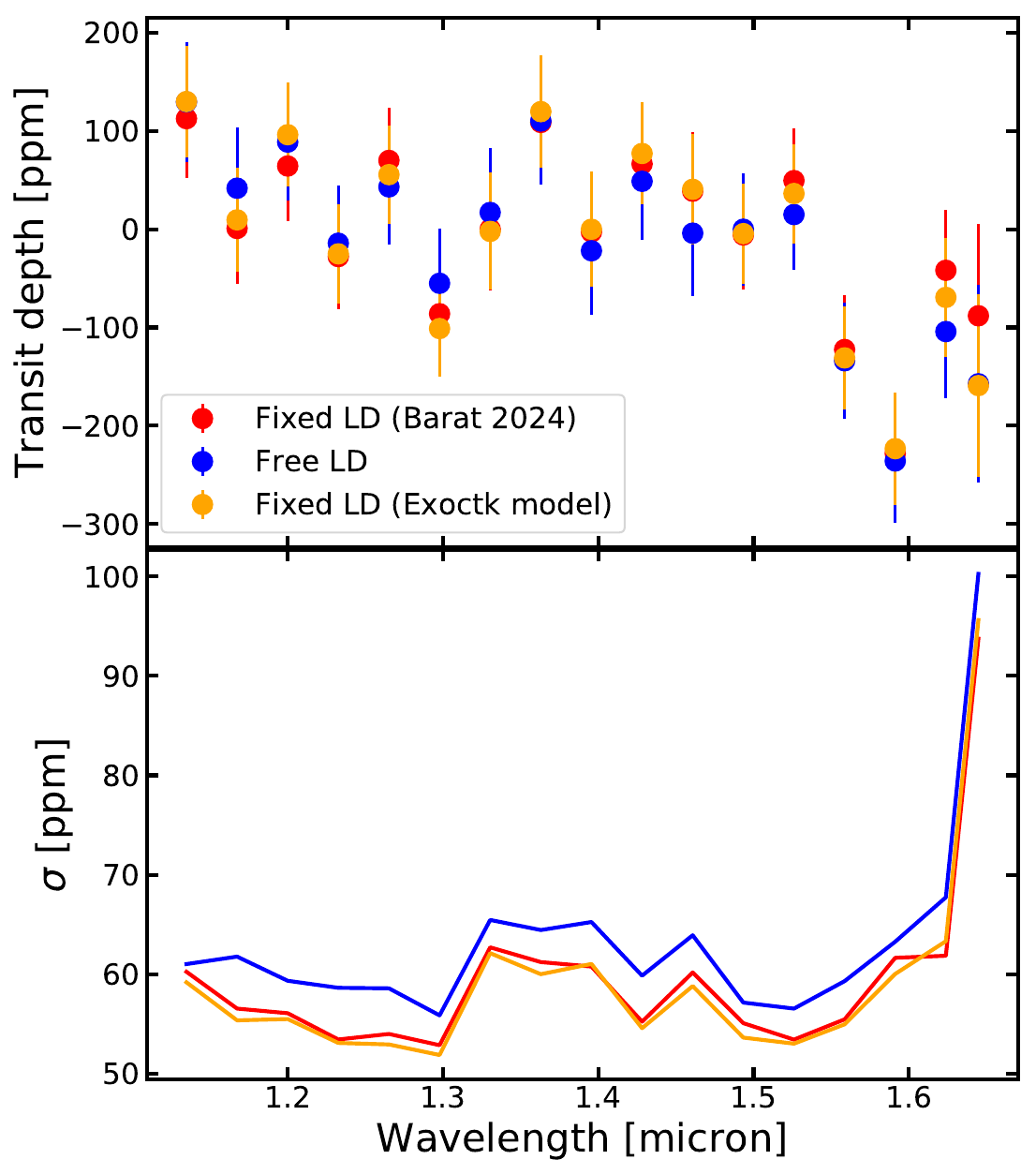}
    \caption{Comparison of the derived transmission spectrum of V1298 Tau c for a fixed and free limb darkening fit to the light curves (upper panel) and corresponding 1$\sigma$ uncertainties for each each spectroscopic bin (lower panel). We fix the limb darkening to i) linear limb darkening coefficients derived from the primary transit of V1298 Tau b \citep{barat2023} shown in red ii) model limb darkening coefficients derived from EXOCTK (Section \ref{v1298_limb_darkening} and Figure \ref{fig:limb_darkening_comparison}) shown in orange. The transmission spectrum derived using free linear limb darkening coefficients for planet c is shown in blue. The median subtracted transmission spectra in all three cases are consistent within 1$\sigma$ of each other for all channels. A comparison between the transit depth uncertainty for the free and fixed limb darkening cases shows that fixing the limb darkening reduces the transit depth uncertainty by 5-10\% for each channel.}

    \label{fig:fix-free-comparison}
\end{figure}

 For the white light curve (1.12-1.6 $\mu$m), we fit the RECTE parameters and the baseline polynomial coefficients, along with the planet's transit depth and semi-major axis. Other orbital parameters were fixed to literature values \citep{Feinstein2022,sikora2023,david2019}. The RECTE parameters constitute 4 parameters: the number of initially filled fast/slow charge traps and the number of fast/slow charge traps that are filled in between orbits. Due to the lack of ingress and egress data, our limb darkening precision was impaired. However, by using the limb darkening parameter retrieved from the observations of planet b \citep{barat2023}, we were able to significantly reduce the uncertainties on the transit spectrum (Figure \ref{fig:fix-free-comparison}). V1298 Tau b and c have impact parameters (0.34$^{+0.19}_{-0.21}$ and 0.46$^{+0.13}_{-0.24}$ respectively) consistent within 1$\sigma$ of each other. Furthermore, \citet{marshall2021} report a relatively low mutual inclination (0$\pm$19$^\circ$), indicating a planar orbital geometry of the V1298 Tau system. Therefore, both planets are likely to transit a similar part of the stellar photosphere with limb darkening during their primary transit. This highlights the advantage and uniqueness of studying multiple planets in the same system. For a discussion on the limb darkening of V1298 Tau see Section \ref{v1298_limb_darkening}. We also independently measure the stellar density from the HST white light curves which is discussed in Section \ref{stellar density}. 

\begin{figure*}
    \centering
    \includegraphics[width=\textwidth]{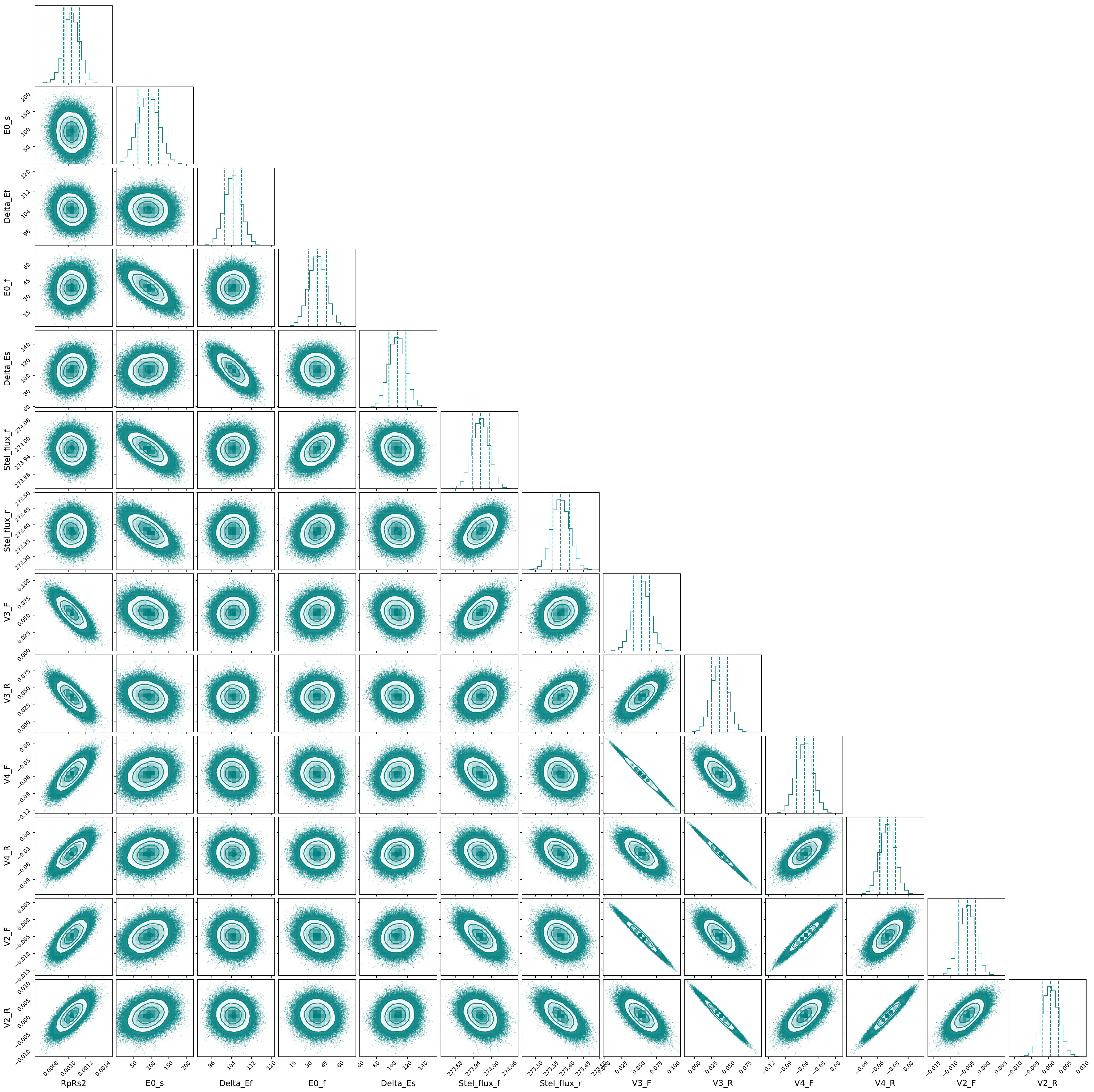}
    \caption{Posterior distribution from the white light curve fit of planet c (See Appendix \ref{light curve fitting}. We use a charge trapping model (RECTE) to model the instrumental systematics \citep{zhou2017} and a second order polynomial to model the out of transit baseline.}
    \label{fig:white_LC_corner}
\end{figure*}

 The fit was performed using \texttt{emcee} \citep{emcee}. The posterior distribution from these fits is shown in Fig. \ref{fig:white_LC_corner} and the best-fit transit model and residuals, are shown in Figure \ref{fig:whitelc} (upper panel).

\begin{figure}
    \centering
    \includegraphics[width=\columnwidth]{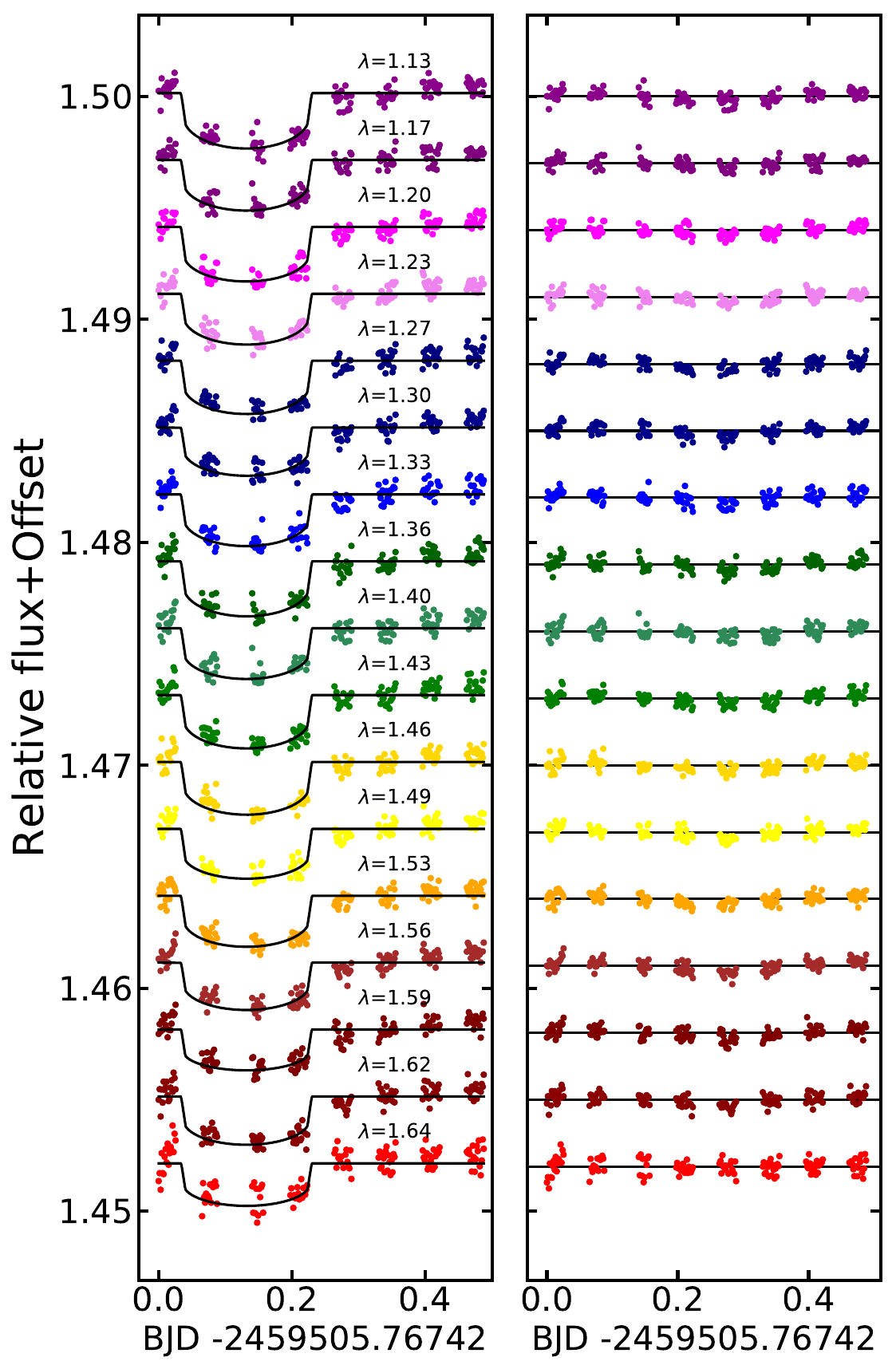}
    \caption{Corrected spectroscopic light
curves (left panel) and residuals (right panel) for the planet V1298 Tau c from HST/WG141 observation. The light-curves
have been offset for visual clarity. The light curve extraction and fitting methodology is described in Appendix \ref{light curve fitting}.}
    \label{fig:spec_lc}
\end{figure}

Spectroscopic light curves were analyzed similarly by dividing the stellar spectra into 7-pixel bins, resulting in 17 channels (identical to V1298 Tau b). We fixed the semi-major axis from the white light curve fits and fit the RECTE parameters and baseline for each bin, and the transit depth. We fix the wavelength dependent limb darkening coefficients to values derived in \citet{barat2023}. The de-trended spectroscopic light curves and residuals are shown in Fig.
\ref{fig:spec_lc}. 

\section{Modeling the baseline} \label{subsection:stellar baseline}

\begin{figure}
    \centering
    \includegraphics[width=\columnwidth]{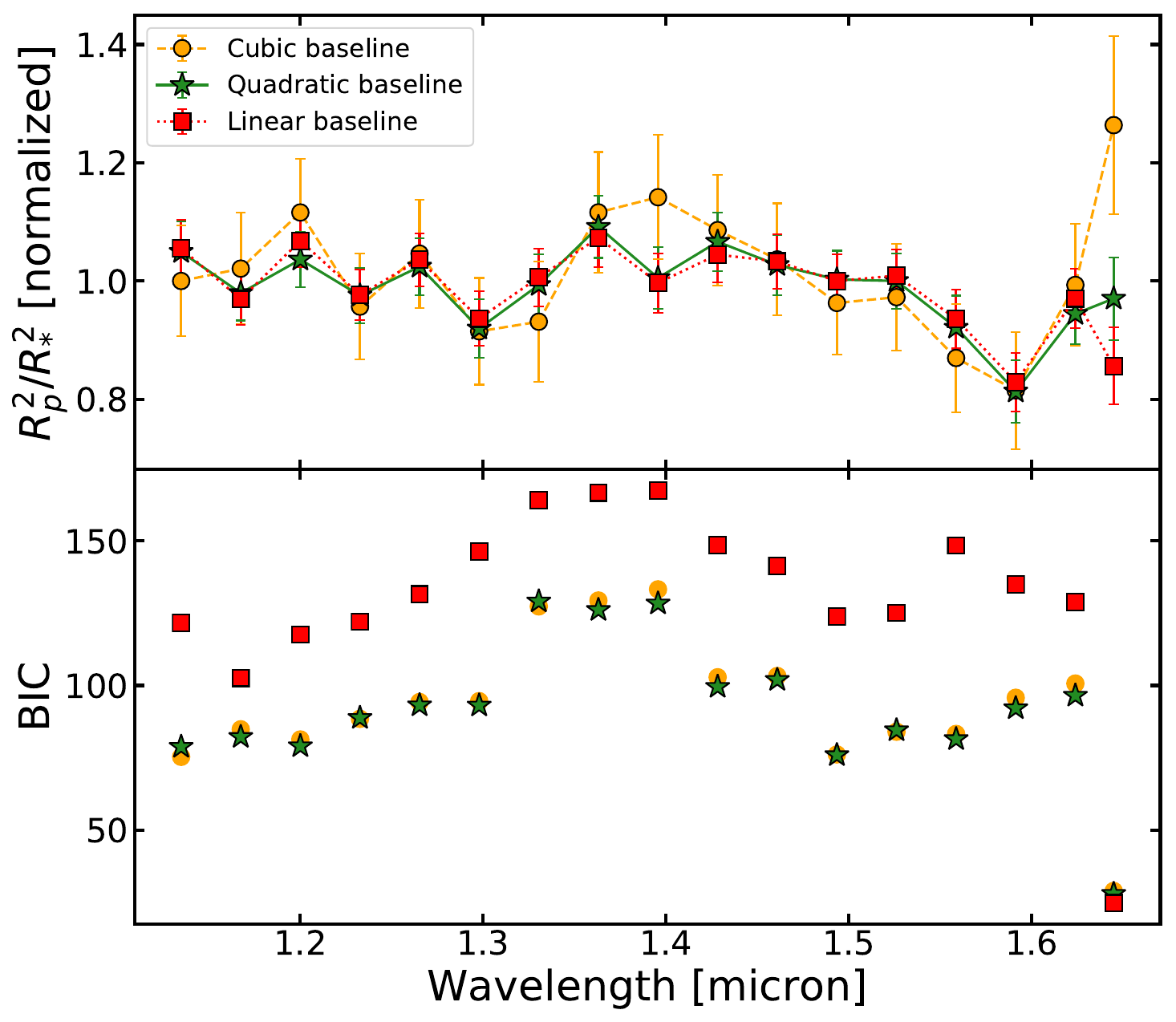}
    \caption{Upper panel: Comparison of transit depths (planet c) for all spectroscopic channels for linear, quadratic and cubic stellar baseline models. The transit depths are normalized by their respective median values for the sake of visual comparison. Lower panel: Comparison of Bayesian Information Criteria (BIC) for linear, quadratic and cubic stellar baseline models. The transit depths agree within 1$\sigma$ of each other for the different polynomial orders, except for the last spectroscopic channel. The BIC clearly favors higher order polynomial models compared to linear for modeling the stellar baseline. For further discussion see Appendix \ref{subsection:stellar baseline}.}
    \label{fig:polynomial comparison}
\end{figure}

Young stars exhibit quasi-periodic flux modulation due to rotational variability \citep[e.g. see][]{david2019b,mann2016,mann2022}. The transit of planet c analyzed in this work occurred near an inflection point of the stellar baseline, resulting in significant deviation from a linear
function. We modelled the baseline using linear, quadratic, and cubic polynomial functions for all the spectroscopic channels. The transit depths from all three fits were consistent within 1$\sigma$, but quadratic and cubic polynomials were statistically favored over a linear baseline (See Fig. \ref{fig:polynomial comparison}).

Higher-order polynomials, however, introduce correlation between the transit depth and baseline, increasing uncertainties
and scatter in the transmission spectrum. The Bayesian Information Criterion (BIC) did not improve significantly with a cubic baseline compared to a quadratic one, and the cubic polynomial coefficients strongly correlated with the transit depths, likely adding noise. Therefore, we used a quadratic polynomial to model the baseline for the final transmission spectrum.

%We assume CH$_4$, HCN and C$_2$H$_2$ as the haze precursor molecules and calculate their abundances from photochemical models. Assuming an efficiency parameter (f$_{haze}$), we derive the radius and density distributions of haze particles. The XUV spectrum of V1298 Tau was adopted from \citet{Duvvuri2023}. We assumed a tholin like composition for the haze \citep{khare1984}. 

\section{Theoretical evolutionary models} \label{appendix: evolution}

The evolutionary models we use in the current study are described in \citet{vazan2022}.
Our models start from the time of disk dispersal and calculate the thermal evolution and structural evolution. We consider both core-envelope as well as gradually mixed interior structures. We include atmospheric mas loss due to photoevaporation from the topmost layer of the planets using the formalism of \citet{rogersandowen2021}. The mass loss rates are estimated from an energy-limited model \citep{erkaev2007}, assuming a fixed photoevaporation efficiency. Our models include core erosion and mixing with the envelope. We choose the initial conditions (mass, radius) such that it matches the current mass and radius for the planets. We assumed 17~M$_\oplus$ (peak of ATMO retrieval posterior distribution) for the current mass estimates of the V1298 Tau c. For planet b, we show two models: 10~M$_{\oplus}$ (consistent within 1$\sigma$ to median mass retrieved from ATMO) and 23~M$_{\oplus}$ (3 $\sigma$ mass upper limit from \citet{barat2023}). We run simulations with both core-envelope and gradually mixed interior models. Both interior structure models can explain the current mass/radius of the planets. Mass and radius (normalized at current value) evolution models for the V1298 Tau planets are shown in Figure \ref{fig:evolutionary models}.

\section{Effect of stellar activity on transmission spectrum} \label{stellar activity}
\begin{figure}
    \centering
    \includegraphics[width=\columnwidth]{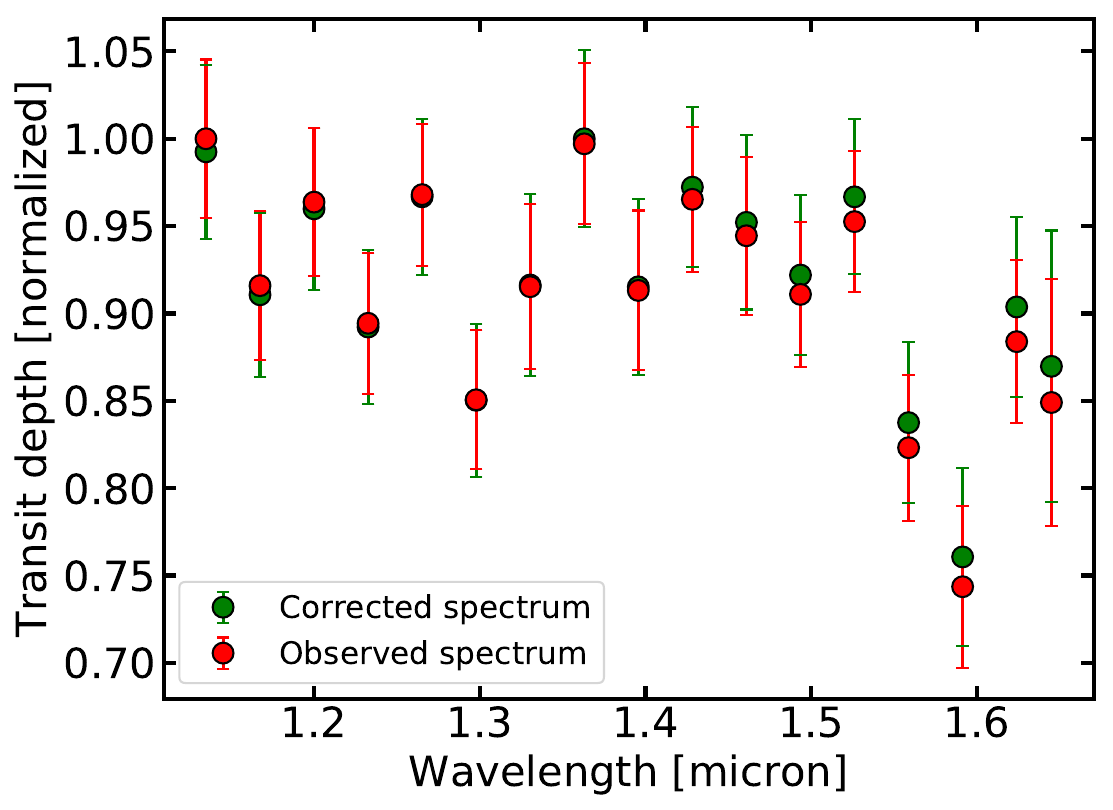}
    \caption{Comparison between observed (red circles) and stellar contamination corrected (green circles) transmission spectrum of V1298 Tau c. The respective spectra have been normalized for the sake of comparison. All the spectroscopic bins are consistent within their 1$\sigma$ uncertainties. The contamination spectra have been derived following the prescription outlined in \cite{rackham2019}, with stellar spot coverage parameters adopted from \cite{feinstein2021}. See Appendix \ref{stellar activity} for further discussion.}
    \label{fig:stellar_corrected_comparison}
\end{figure}

We estimate the effect of stellar activity on the transmission spectrum of V1298 Tau c. We adopt the method outlined in \cite{rackham2019}, where we estimate a correction factor based on the spot coverage fraction and spot temperature contrast. We adopt a spot coverage fraction of 20\% for V1298 Tau \citep{feinstein2021}. We adopt an extreme spot contrast of 1000~K following spot contrasts derived for a sample of T Tauri stars \citep{koen2016}. We show a comparison between the observed and corrected transmission spectrum in Fig. \ref{fig:stellar_corrected_comparison}. The observed and contamination corrected normalized spectra are consistent within 1$\sigma$ for all spectroscopic channels.

Since V1298 Tau is a K-type star, its contamination function does not show molecular features like M-dwarfs \citep{barclay2021}. Therefore, the main effect of the stellar correction is to change the level of the continuum in this case. However, the atmospheric features, which are relative to the continuum remain largely unaffected.

\section{Supplementary Figures}

\begin{figure}
    \centering
    \includegraphics[width=1\linewidth]{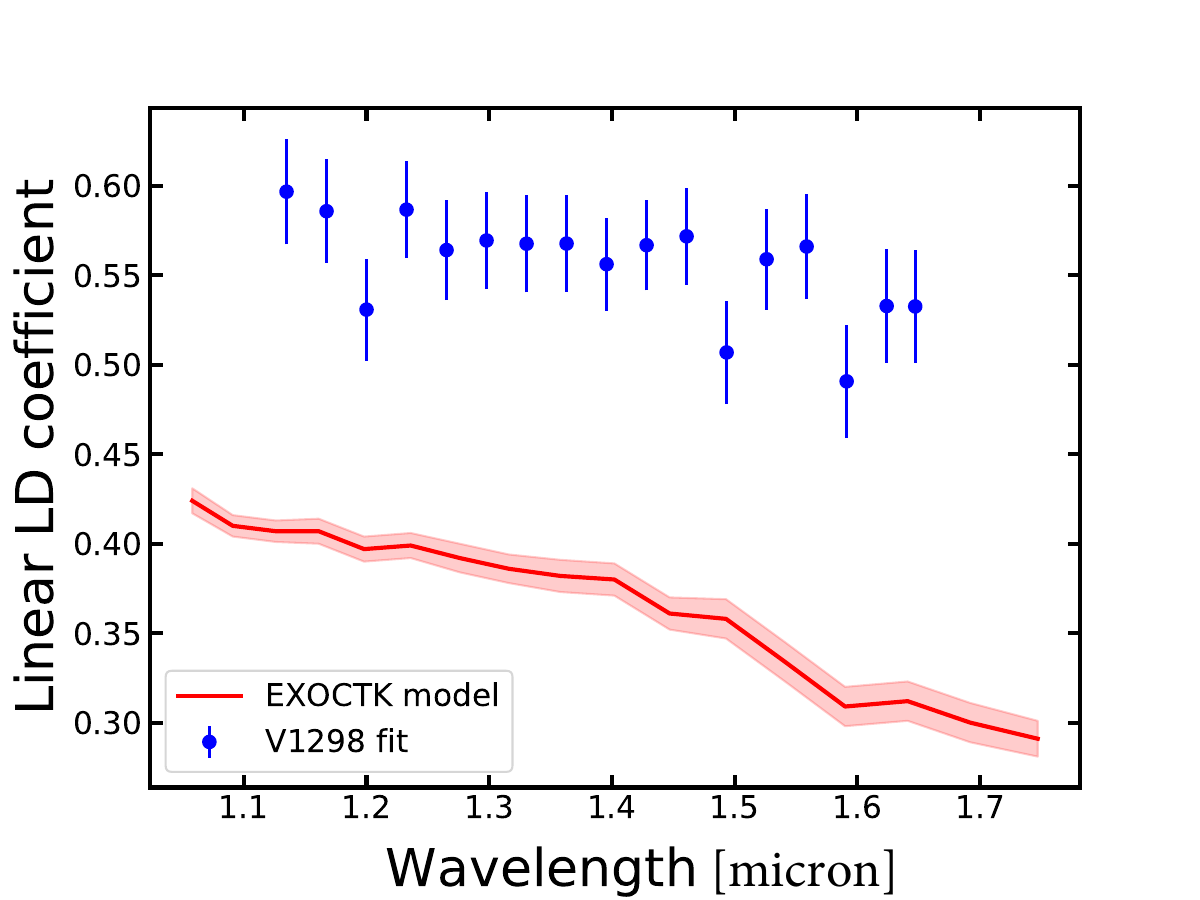}
    \caption{Comparison between the fitted linear limb darkening coefficients derived from the V1298 Tau b visit \citep{barat2023} and theoretical limb darkening model for a main sequence star of a similar spectral type. The theoretical models have been derived using EXOCTK \citep{bourque2021} for a star with similar spectral type as V1298 Tau (K1). We used photospheric temperature of 5000K, log~g=4 and solar metallicity \citep{finoceti2023} PHOENIX models to simulate the theoretical linear limb darkening coefficients in the HST/WFC3 G141 bandpass. To our knowledge, this is the first time that the limb darkening of a weak-lined T-Tauri star has been measured. The measured limb darkening appears significantly different compared to that of a main sequence star of the same spectral type. This discrepancy between with main-sequence models is could be due to different interior structure and high magnetic field of this young star (See Section \ref{v1298_limb_darkening}). This figure highlights the importance of fitting for the limb darkening when analyzing transit light curves around such young stars.}
    \label{fig:limb_darkening_comparison}
\end{figure}

\begin{figure}
    \centering
    \includegraphics[width=1\linewidth]{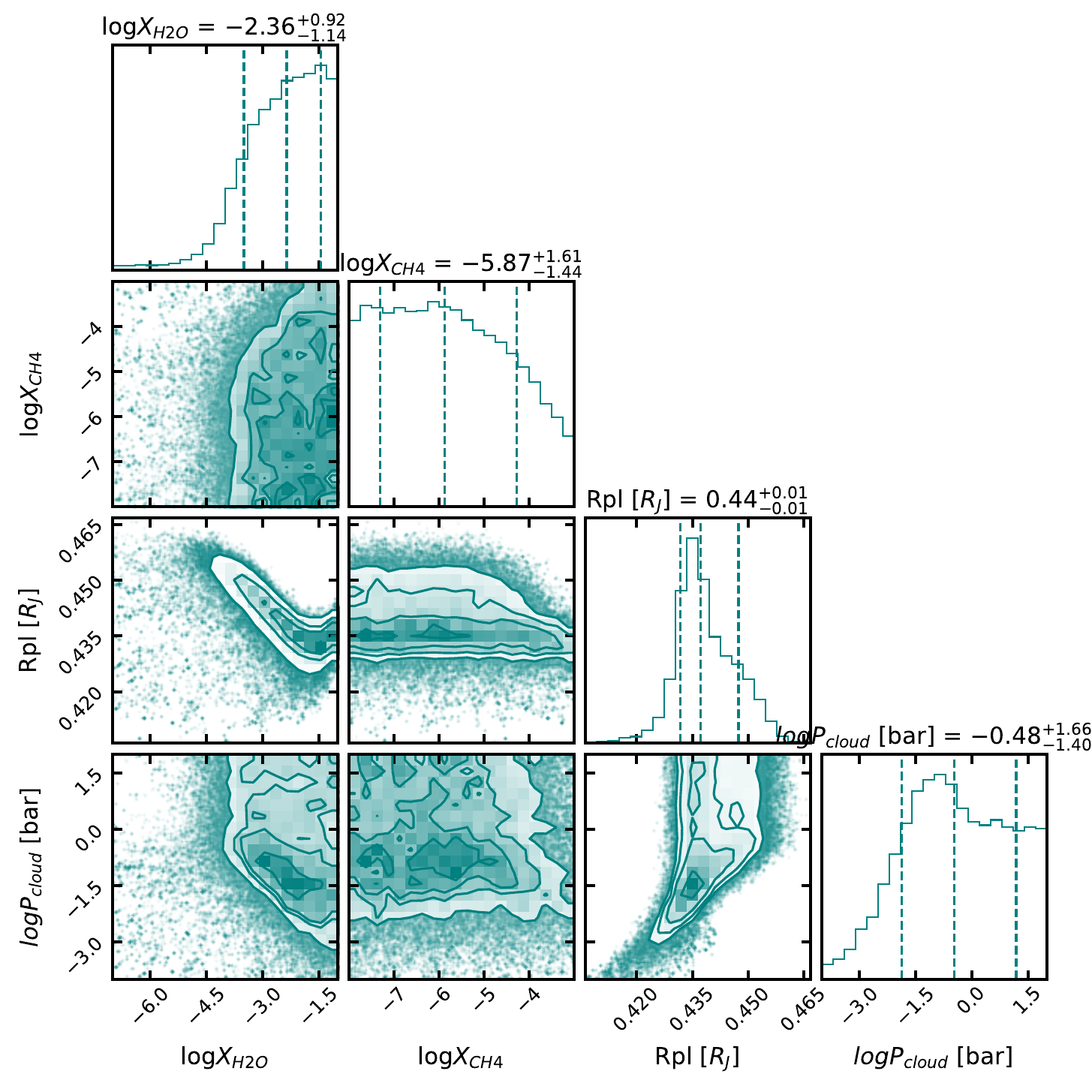}
    \caption{Posterior distribution from free atmospheric chemistry retrieval on V1298 Tau c using \texttt{PetitRadtrans}. We include H2O and CH4 opacity in our forward model. We fixed the mass 17~M$_{\oplus}$ and the isothermal temperature to 950~K. The results from this free retrieval are discussed in Section \ref{subsec:transmission spectrum} }
    \label{fig:free corner}
\end{figure}

\begin{figure}
    \centering
    \includegraphics[width=1\linewidth]{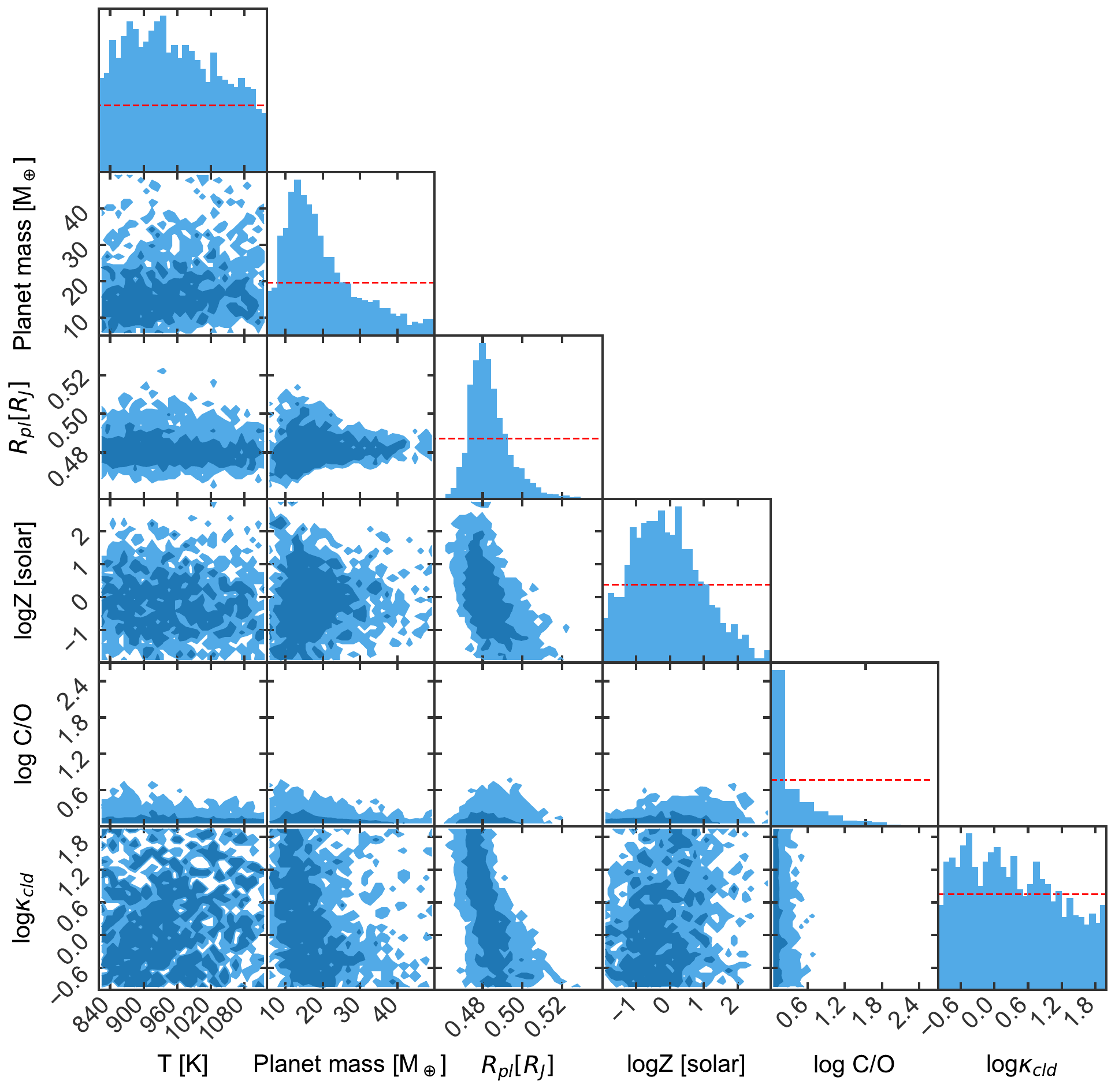}
    \caption{Posterior distribution from retrievals on V1298 Tau c transmission spectrum using ATMO equilibrium chemistry models where planet mass has been included as a free parameter. The red horizontal lines show the prior distributions used (See Section \ref{subsec:atmo}). The different shades on the correlation plots show 1 and 2$\sigma$ confidence regions.   The retrieved parameters are tabulated in Table \ref{tab:table1}. Description of models used for this fit are given in Section \ref{subsec:atmo}.}
    \label{fig:atmo_posterior}
\end{figure}

\begin{figure}
    \centering
    \includegraphics[width=1\linewidth]{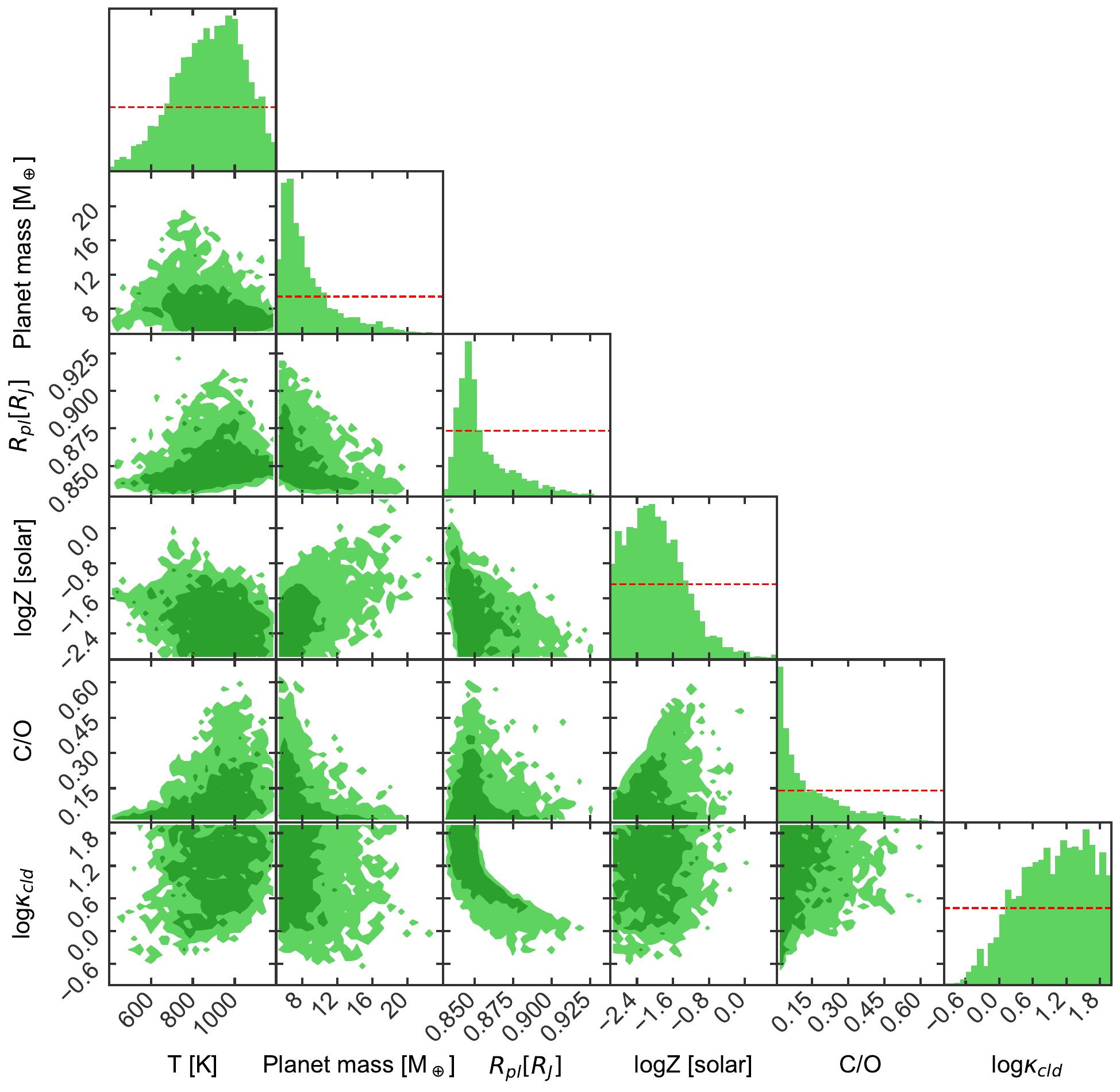}
    \caption{Same as previous figure for planet c.}
    \label{fig:atmo_posterior_b}
\end{figure}

\begin{figure}
    \centering
    \includegraphics[width=1\linewidth]{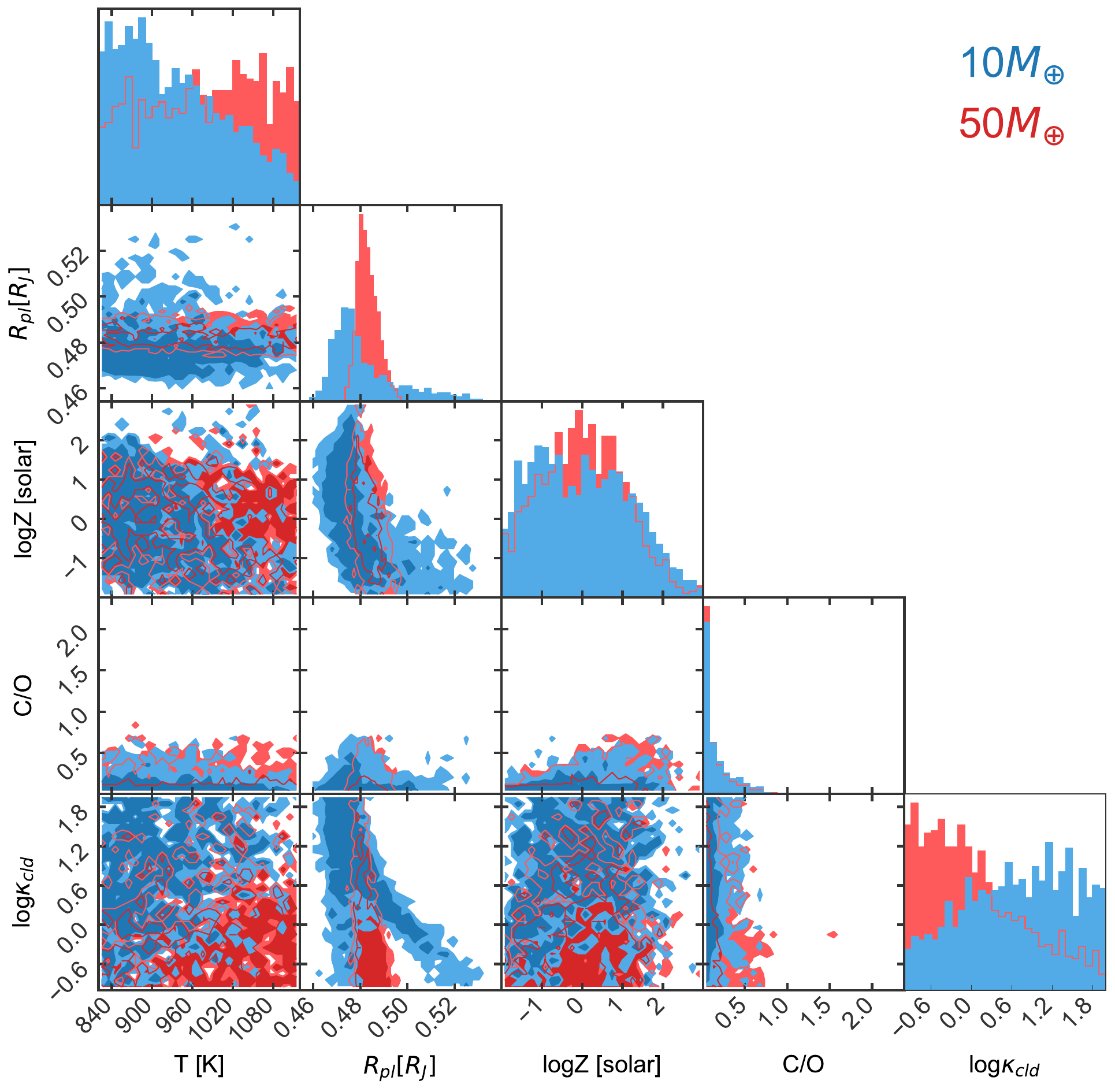}
    \caption{ Posterior distribution from ATMO retrievals of the transmission spectrum of V1298 Tau c by assuming fixed planet mass of 10~M$_\oplus$ (blue) and 50~M$_\oplus$ (red). The median values of the fitted parameters are shown in Table \ref{tab:table1}. The priors used for these fits are same as those shown in Figure \ref{fig:atmo_posterior}. Description of models used for this fit are given in Section \ref{subsec:atmo} }
    \label{fig:10_50_comparison_corner}
\end{figure}

\begin{figure}
    \centering
    \includegraphics[width=1\linewidth]{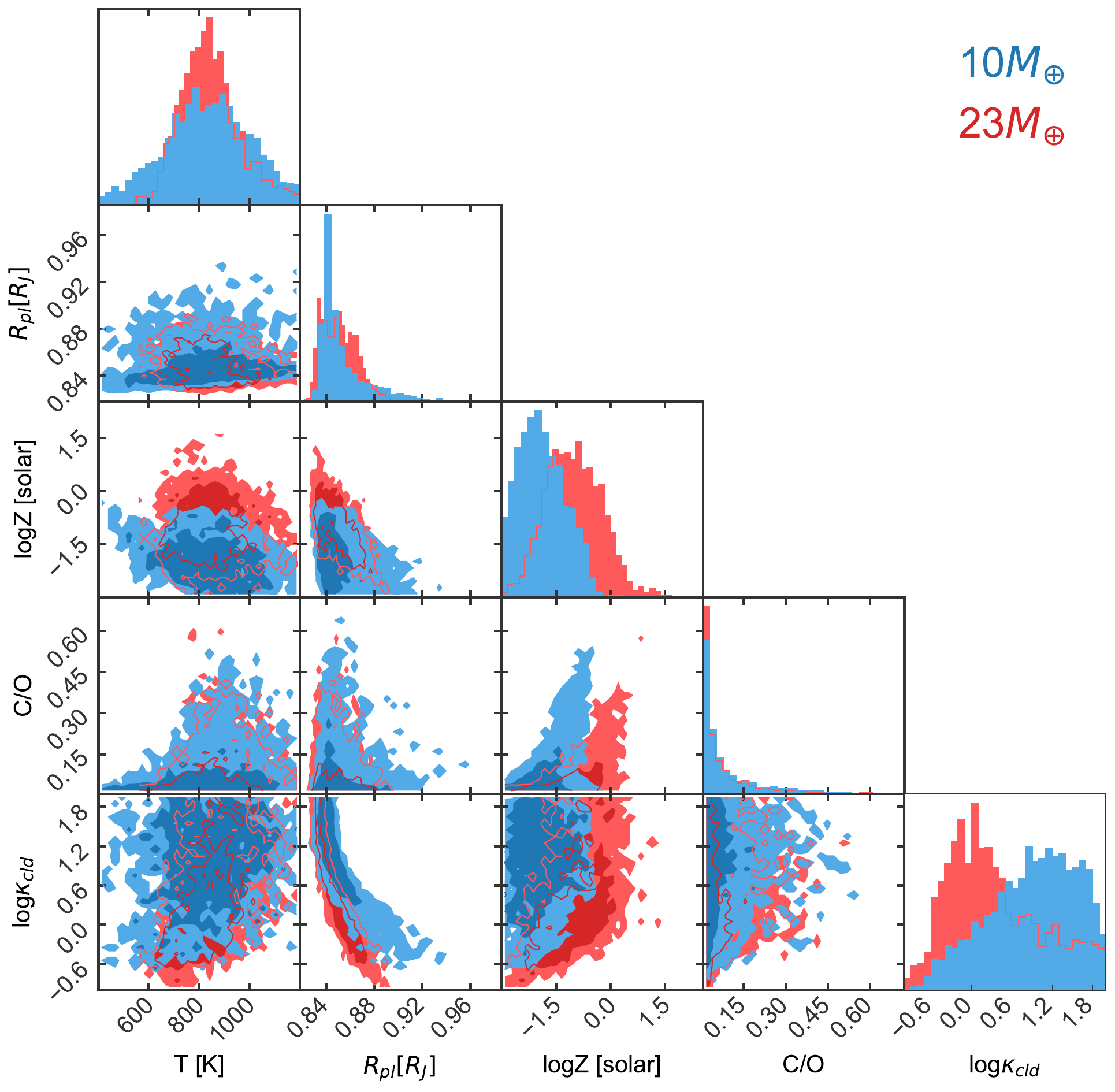}
    \caption{ Posterior distribution from ATMO retrievals of the transmission spectrum of V1298 Tau b by assuming fixed planet mass of 10~M$_{\oplus}$ (blue) and 23~M$_{\oplus}$ (red). The median values of the fitted parameters are shown in Table \ref{tab:table1}. The priors used for these fits are same as those shown in Figure \ref{fig:atmo_posterior_b}. Description of models used for this fit are given in Section \ref{subsec:atmo} }
    \label{fig:10_23_comparison_corner}
\end{figure}

\newpage

\begin{table*}
    \centering
    \caption{Mass and atmospheric metallicities of planets with measured metallicity using JWST which have been shown in Figure \ref{fig:metallicity-distance}}
    \resizebox{\textwidth}{!}{\begin{tabular}{c | c | c |c |c }
    \hline
        Planet name & Mass [M$_{J}$] & Atmospheric metallicity [solar] & Semi-major axis [AU]  & Stellar metallicity [dex]   \\
        \hline
        WASP-18b \tablefootmark{1} &	10.20$\pm$0.35  &   1.03$^{+1.1}_{-0.51}$ & 0.02 & 0.11$\pm$0.08   \\

        WASP-39b \tablefootmark{2,3,4,5} &  0.280$\pm$0.03  &    10 & 0.048 & -0.03$\pm$0.1 \\

        WASP-77Ab \tablefootmark{6} &	1.66$\pm$0.06 &	0.12$^{+0.1}_{-0.05}$ & 0.023 & -0.01$\pm$0.1  \\

        WASP-96b \tablefootmark{7} &	0.48$\pm$0.03 &	0.23$^{+0.7}_{-0.1}$ &    0.045 & 0.14$\pm$0.19  \\

        TOI-270d \tablefootmark{8}	& 0.015$\pm$0.001	& 225$^{+86}_{-98}$ &	0.07 & -0.2$\pm$0.1  \\

        WASP-107b \tablefootmark{9} &	0.1$\pm$0.005 &	43$^{+8}_{-8}$ & 0.05 & 0.02$\pm$0.09  \\

        HD149026b \tablefootmark{10} & 0.358$\pm$0.01 & 	 20$^{+11}_{-8}$ &   0.0436 & 0.33$\pm$0.1  \\

        HD209458b \tablefootmark{11} &	0.5902904$\pm$0.01 &	3$^{+4}_{-1}$ &	0.05 & -0.01$\pm$0.05   \\

        WASP-80b \tablefootmark{12} &	0.53$\pm$0.03 &	6$^{+4}_{3}$ &     0.03 & -0.1$\pm$0.1  \\

        WASP-17b \tablefootmark{13}	& 0.78$\pm$0.2   &  50  &  	0.051 & -0.07$\pm$0.1  \\

        GJ1214b \tablefootmark{14}  &	0.026$\pm$0.001  & $>$100 &       0.01  & 0.02$\pm$0.1  \\
       \hline
       \hline
    \end{tabular}}
 %   \caption{Mass and atmospheric metallicities of planets with measured metallicity using JWST. In Figure \ref{fig:metallicity-distance} we show these planets in the mass-metallicity and metallicity-distance diagram to compare with the V1298 Tau system. The stellar metallicity provided in this table has been used to convert the derived atmospheric metallicity from atmospheric  in solar units to stellar units. For WASP 39b we quote 10$\times$ solar metallicity, without any errorbars, since this value is common between all the four JWST ERS papers where metallicity ranges are estimated from grid comparisons. Similarly, for WASP 17b, the best fit metallicity is derived from grids, and do not provide uncertainties. }
 \tablefoot{The stellar metallicity provided in this table has been used to convert the derived atmospheric metallicity from atmospheric  in solar units to stellar units. For WASP 39b we quote 10$\times$ solar metallicity, without any errorbars, since this value is common between all the four JWST ERS papers where metallicity ranges are estimated from grid comparisons. Similarly, for WASP 17b, the best fit metallicity is derived from grids, and do not provide uncertainties.\\
 \tablefoottext{1}{\citet{coulomb2023}}
 \tablefoottext{2,3,4,5}{\citet{rustomkulov2023,feinstein_ers,ahrer_ers,alderson_ers}}
 \tablefoottext{6}{\citet{August2023}}
\tablefoottext{7}{\citet{radica2023}}
\tablefoottext{8}{\citet{benneke2024}}
\tablefoottext{9}{\citet{sing2024}}
\tablefoottext{10}{\citet{gagnebin2024}}
\tablefoottext{11}{\citet{xue2024}}
\tablefoottext{12}{\citet{bell2023}}
\tablefoottext{13}{\citet{Grant2023}}
\tablefoottext{14}{\citet{kempton2023}}

 }
    \label{tab:jwst_metallicity_table}
\end{table*}

\end{document}